# Room-scale CO$_2$ injections in a physical reservoir model with faults

M.A. Fernø[1,4], M. Haugen[1], K. Eikehaug[1], O. Folkvord[1], B. Benali[1], J.W Both[2], E. Storvik[2], C.W. Nixon[3], R.L. Gawthrope[3] and J.M. Nordbotten[2,4]

[1] Dept. of Physics and Technology, University of Bergen
[2] Center for Modeling of Coupled Subsurface Dynamics, Dept. of Mathematics, University of Bergen
[3] Dept. of Earth Science, University of Bergen
[4] Norwegian Research Center, Postboks 22 Nygårdstangen, 5838 Bergen

Abstract

We perform a series of repeated CO$_2$ injections in a room-scale physical model of a faulted geological cross-section. Relevant parameters for subsurface carbon sequestration, including multiphase flows, capillary CO$_2$ trapping, dissolution, and convective mixing, are studied and quantified. As part of a forecasting benchmark study, we address and quantify six predefined metrics for storage capacity and security in typical CO$_2$ storage operations. Using the same geometry, we investigate the degree of reproducibility of five repeated experimental runs. Our analysis focuses on physical variations of the spatial distribution of mobile and dissolved CO$_2$, multiphase flow patterns, development in mass of the aqueous and gaseous phases, gravitational fingers, and leakage dynamics. We observe very good reproducibility in homogenous regions with up to 97 % overlap between repeated runs, and that fault-related heterogeneity tends to decrease reproducibility. Notably, we observe an oscillating anticline CO$_2$ leakage behavior from an open anticline with a spill point in the immediate footwall of a normal fault, and discuss the underlying causes for the observed phenomenon within the constraints of the studied system.

## 1. Introduction

In its simplest form, carbon sequestration involves the injection of captured carbon dioxide (CO$_2$) into deep subsurface porous and permeable sedimentary rocks, overlain by an impermeable sealing layer. The migration of the buoyancy-driven CO$_2$ is determined by: i. the intrinsic rock and fluid properties (e.g. porosity, permeability, fluid density, and viscosity); and ii. the distribution and properties of geological structures such as faults and fracture networks, that are inherent to both reservoir and seal rocks. Faults are discontinuities that form at a range of scales; they can act as conduits or barriers for flow, and they generally have directionally-dependent flow properties [Bastesen and Rotevatn, 2012]. Large sealing faults control storage site geometries and compartmentalization, whereas networks of small faults and fractures may affect reservoir flow and seal integrity [Ogata et al, 2014].

*Faults, fractures and flow*

The properties of the fracture networks (i.e. topology, connectivity, permeability) that form damage zones around faults as they evolve [Nixon et al, 2020] are particularly important to CO$_2$ flow. Subsurface faults are discerned from reflection seismic data, but descriptions suffer from limitations in seismic resolution and coverage. Geologically analogous outcrops and dedicated laboratory experiments provide a means to investigate smaller structures around faults and shed light on flow and sealing properties. Being able to identify and forecast the behavior of potential subsurface bypass structures during CO$_2$ injection is essential; understanding the interplay between multiphase flow and fault evolution is critically needed for carbon sequestration projects. Despite this, the flow properties of faults and their damage zones remain insufficiently understood, and little is known about how their flow behavior evolves in the different stages of a carbon sequestration project. Our current understanding of large-scale CO$_2$ plume migration is mainly from time-lapse seismic surveys with limited a priori knowledge [Furre et al, 2017]. With increases in reservoir pressures during CO$_2$ injection, there is a greater risk of reactivation and potential generation of new fracture networks that can enhance seal permeability, capillary flow and provide pathways for fluid escape to shallower reservoirs or the surface (e.g. Ogata et al., 2014; Karstens and Berndt, 2015; Karstens et al., 2017).

*Forecasting skills*

Accurate modeling and simulation of multiphase flow in porous media with faults is central to carbon sequestration forecasting, risk assessment and mitigation strategies. Geologically accurate models are needed for simulating flow responses of faults, where analogous outcrops and laboratory experiments may be used to discern spatial variations within faults [Rotevatn et al, 2009] to determine percolation potential. Susceptibility for reactivation upon pressurization can then be evaluated with geomechanical modelling. Forecasting of field-scale $CO_2$ migration and reservoir pressure response is commonly achieved with history-matching and extrapolation exercises, routinely applied in hydrocarbon production. This approach can be effective with a limited number of parameters [Sharma et al, 2019], but flow simulations that include geologically realistic descriptions of faults and fracture networks are computationally expensive in universally applied industrial simulators. Current field-scale simulation approaches requires significant approximations, with respect either to spatial and temporal resolution, or to coupled processes, or both. Therefore, insights into scale-dependent flow behavior are needed to better couple flow dynamics in the presence of fault-related heterogeneity.

*The laboratory FluidFlower rig and its relevance to subsurface storage*

The FluidFlower concept links research and dissemination through a new experimental rig constructed at University of Bergen (UiB) that enables meter-scale, multiphase, quasi two-dimensional flow on model geological geometries with unprecedented data acquisition and repeatability. The geological geometry of the physical room-scale model (cf. **Figure 1**) is motivated by typical North Sea reservoirs. Structurally, the benchmark geometry is characterized by broad open folds and normal faults: a major normal fault breaches the lower reservoir-seal system and terminates upward at the base of the upper reservoir. A broad open anticline, in the footwall of the fault, forms the main trap to the lower reservoir-seal system and has a spill point in the immediate footwall of the fault. The broad open anticline is also the main trap geometry for the upper reservoir-seal system, but this is affected by a graben bounded by two oppositely dipping normal faults. Hence, the designed geometry focuses on the need to couple fault and flow behavior during $CO_2$ injection, and was designed to achieve realistic $CO_2$ flow and trapping mechanisms to benchmark the numerical modelling capability of the porous media community with new physical measurement of key processes.

While the present study is at the laboratory scale, the fundamental physical processes of multiphase, multi-component flows in heterogeneous porous media are the same as at reservoir conditions. The most important subsurface $CO_2$ trapping mechanisms are present in the FluidFlower rig: *structural trapping* occurs under the sealing sand layers and within different reservoir zones; *dissolution trapping* occurs almost instantaneously when the injected $CO_2$ dissolves into the water phase initially saturating the porous media; *residual trapping* is observed in regions with intermediate water saturation, but is temporary because of rapid dissolution; *convective mixing* occurs when the $CO_2$-saturated water migrates downwards and generate gravitational fingers. *Mineral trapping* is by design not part of the benchmark study for increased control of active chemistry (very pure silica sand is used, and the pressure and temperature conditions are set outside mineralization thresholds within the experimental time series). Hence, the observed flow and trapping behavior in the FluidFlower rig to a large degree represents the physical in a subsurface system, even if the petrophysical properties like porosity and permeability, as well as the pressure and temperature conditions are not directly comparable to subsurface conditions. Furthermore, we remark that the structural trapping in the FluidFlower relies more on capillary entry pressure, and less on permeability contrast, than expected at the field scale. Overall, we argue that the findings and observations in this study are indicative of field-scale simulation, although several observed phenomena scale differently in the FluidFlower compared with subsurface systems (for a detailed scaling analysis, see Kovscek et al 2023, this issue).

Despite the physical similarities, actual field scale simulation will deviate from this study in several important aspects, of which we highlight (see Flemish et al 2023, this issue, for a comprehensive discussion):
- *Heterogeneity.* The facies in the benchmark geometry were built with a single sand type aiming for homogenous petrophysical properties, and, hence, emphasizing larger scale structural heterogeneities. On the field scale, it is expected that there will be significant subscale heterogeneity also within each geological structure.
- *Quality of geological characterization.* A high-resolution image of the geological geometry, with accompanying thicknesses before $CO_2$ injections, was issued to the benchmark participants (cf. Nordbotten et al 2022). At the field scale, the initial geological characterization will associated with higher uncertainty and lower spatial resolution data from seismic surveys.
- *Pressure and temperature conditions.* The laboratory conditions in the reported study yield a gaseous $CO_2$ phase when injected, compared with liquid or supercritical phase at field conditions in typically reservoirs. The difference in phase condition has a minor impact on viscosity, but leads to a denser and less compressible $CO_2$ phase at the field scale.

The importance of forecasting, risk assessment and mitigation strategies for carbon sequestration, with many of the critical coupled subsurface processes remaining poorly understood, merits a continued broad interdisciplinary engagement. The utility of numerical modeling and simulation as a key decision-making tool for industrial application of $CO_2$ storage is scrutinized in the FluidFlower international benchmark initiative.

## 2. Materials and Methods

This section briefly describes the key operational considerations and methodology developed to perform the experimental part of the Benchmark study. It provides an overview of all procedural steps, a description of the geological geometry and parameters. The description is not exhaustive, and the reader is referred supplementary materials (SM) and cited work for more detailed descriptions.

**2.1 Fluids**

The main fluids and their composition and usage are listed in Table 1.

**Table 1**: Fluid compositions and role in benchmark study

| Fluid | Phase | Composition | Usage |
|---|---|---|---|
| pH sensitive solution, termed 'formation water' | Aqueous | deionized water with<br>- 0.14 mM bromothymol blue ($BTB^-$)<br>- 0.43 mM methyl red ($MRe^-$)<br>- 0.10 mM hydroxide ($OH^{-1}$)<br>- 0.67 mM sodium ions ($Na^+$) | Saturate the pore space and enable detecting of dissolved $CO_2$ in the aqueous phase |
| $CO_2$ | Gaseous | 99.999% - 5.0 purity | Injected as gaseous phase |
| Lye solution | Aqueous | Deionized water with<br>0.48 mM sodium hydroxide (NaOH) | Cleaning fluid to remove $CO_2$-saturated water and trapped free gas. |
| Acid | Aqueous | Tap water with > 0.1 M hydrochloric acid (HCl) and pH < 1.0 | Sand cleaning |

Throughout the article we refer to the gaseous form of $CO_2$ as '*gas*' – the dry gas injected will partially partition into the aqueous phase saturating the porous media, and will have a positive, non-zero water content due to solubility of water in $CO_2$. The water content in $CO_2$ was not explicitly quantified in this work. We refer to the aqueous phase partially saturated with dissolved $CO_2$ as the '*$CO_2$-saturated water*', and the aqueous phase without $CO_2$ as '*formation water*'. The aqueous, pH-sensitive solution ('formation water') was in equilibrium with the atmosphere when injected and contained dissolved atmospheric gases (predominantly nitrogen and oxygen). The presence of other gases influences the $CO_2$-to-water mass transfer due to differences in gas-to-water Henry's constant [Van De Ven and Mumford, 2020]: the $CO_2$ mass transfer to the formation water releases nitrogen and oxygen into the gaseous phase. Hence, over time the gaseous phase in the system becomes deprived of $CO_2$, with reduced solubility in water. This effect was predominately observed towards the later-life of the gas accumulation under the anticlines, and is discussed more below.

## 2.2 Sand handling and porous media flow properties

Danish quartz sand was purchased (in total 3.5 tons) and systematically treated to achieve the required properties. Six different sand types were used in the benchmark study (see **Table 2**). Before use, each sand was manually sieved from the supplied sand stock and treated with a strong acid (HCL) to remove impurities (predominately calcite). The acid was neutralized with sodium hydroxide, rinsed with tap water while manually agitating to remove precipitates and dust until no visible particles, then rinsed in tap water multiple times until clear solution without particles. The sand was then dried at 60 °C until dry and stored in cleaned plastic containers with lid until use. The absolute permeability was measured for each sand, all with nominal porosity 0.44. Detailed sand description, properties and procedural steps are outlined in [Haugen et al. 2023, this issue].

**Table 2**: Key parameters for each of the six sand types

|  | Grade | <grain size>[a] ± σ [mm] | Nominal K [D] | $P_{C\_entry}$[b] [mbar] |
|---|---|---|---|---|
| **Sand ESF** | Fine | 0.20 ± 0.11 | 50 | 15.0 |
| **Sand C** | Coarse (lower) | 0.66 ± 0.09 | 500 | 3.3 |
| **Sand D** | Coarse (upper) | 1.05 ± 0.14 | 1 000 | 0.9 |
| **Sand E** | Very coarse (lower) | 1.45 ± 0.19 | 2 000 | *0.26* |
| **Sand F** | Very coarse (upper) | 1.77 ± 0.31 | 4 000 | *0.10* |
| **Sand G** | Granules | 2.51 ± 0.63 | 10 000 | *0.01* |

[a] averaged smallest grain width reported for each sand. Grains are not circular.

[b] Capillary entry pressures measured from gas column height (in mm) sustained under each sand and converted to mbar. Italic numbers extrapolated from trend, no observable gas column.

## 2.3 The FluidFlower rig and building the geometry

The *FluidFlower* enables meter-scale multiphase quasi two-dimensional flow experiments on model geological geometries with unprecedented data acquisition and repeatability. Time-lapsed images are acquired to monitor dynamic, multiphase flow patterns with high spatial resolution where single sand grains may be identified. The $CO_2$-saturated water is distinguished from formation water by a color shift of aqueous pH sensitive solution, whereas the gas phase is observed by reduction in colored aqueous phase (formation or $CO_2$-saturated water). The design allows for repeated injections tests with near identical initial conditions, allowing physical uncertainty and variability to be addressed using the same geological geometry. The model geological geometry is constructed using unconsolidated sands (cf. **Table 2**) and held in place between an optically transparent front panel and an opaque back panel. The rig has 56 perforations that enable a range of well configurations (injector, producer, monitoring, or plugged) for porous media flow studies.

The FluidFlower rig is curved to sustain internal forces and capable of porous media up to approximately 6 $m^2$ (3 m length x 2 m height). The benchmark study monitored four wells (two for $CO_2$ injection and two for pressure measurements), but several other wells were active during the experiments. Technical wells/ports at the bottom and top enabled re-setting fluids between $CO_2$ injections and to maintain a fixed water column experiments. Technical considerations and mechanical properties of the FluidFlower rig are detailed elsewhere [Eikehaug et al 2023a, this issue]. The FluidFlower has no-flow boundaries at the bottom and both sides, whereas the top is open with a fixed free water column (constant hydrostatic head). Relevance for subsurface carbon sequestration processes is maintained as dominant multiphase flow parameters and trapping mechanisms are present in the room-scale laboratory flow rig, including capillarity, dissolution, and convective mixing.

The dry, unconsolidated sands were meticulously poured from the top into the water-filled void between the front and back panels. Each layer (consisting of one sand type, except the heterogeneous fault) was constructed from the bottom and upwards, and faults and large dipping angle were created by manipulating the layer during pouring using guiding polycarbonate rectangles, funnels and plastic hoses. Mechanical manipulation (raking/scratching) was kept to a minimum and only in some areas in the vicinity of the faults. Faults were constructed though an iterative process, detailed in [Haugen et al 2023, this issue], and the sealed fault was created using a silicone rubber rectangle. The hydrostatic pressure during geometry assembly was 100 mm above operating conditions. When the geometry was complete, the water-level was lowered to operating water-level (kept constant during all injections). Multiple flushing sequences using injection rates 10% higher than the injection protocols (cf. **SM 4**) were performed to achieve an initial, pre-injection sand settling to improve conditions for reproducibility during $CO_2$ injections. The nominal porous media depth was 19 mm, but depth variations were observed and accounted for with a spatially resolved depth map (cf. **SM 2**).

**2.4 The rationale behind the Benchmark geometry**
The geological geometry of the physical room-scale model (cf. **Figure 1**) was motivated by typical North Sea reservoirs. It was developed in close interdisciplinary collaboration between UiB researchers from reservoir physics, earth science and applied mathematics based on the following four principles:

1. Incorporate relevant features frequently encountered in subsurface geological carbon sequestration
2. Enable realistic $CO_2$ flow patterns and trapping scenarios with increasing modeling complexity
3. Sufficiently idealized for the sand facies to be reproduced numerically with high accuracy
4. Be able to operate, monitor and reset the fluids within a reasonable time frame

The geometry was designed to achieve realistic $CO_2$ flow and trapping mechanisms to evaluate the modelling capability of the porous media community. The anticipated $CO_2$ flow, migration and phase behavior from each of the two $CO_2$ injection wells are described below, along with a geological interpretation of the benchmark geometry where geological features described are found in **Figure 1** and highlighted in *italic* below.

*Geological interpretation of benchmark geometry*
The benchmark geometry is a compromise between geological realism, building a physical model from unconsolidated sand, and accurate gridding for numerical simulations of the geometry. The benchmark geometry comprises two stacked reservoir-seal systems, each capped by regional seals (represented by sand ESF). The lower reservoir is a homogeneous, high permeability reservoir (sand F) overlain by a laterally continuous seal. In contrast, the upper reservoir is stratigraphically more heterogeneous, forming an overall upward fining succession, but with permeability variations within the coarse sand layers (alternation of sands E, F, D and C), and additional stratigraphic complexity around a *sealed fault* associated with the local development of sands C and D.

Structurally, the benchmark geometry is relatively simple, characterized by broad open folds and normal faults. The major left-dipping normal fault (*heterogeneous fault*) breaches the lower reservoir-seal system and terminates upward at the base of the upper reservoir (within sand F). A broad open anticline, in the footwall of the fault, forms the main trap to the lower reservoir-seal system and has a spill point in the immediate footwall of the fault. The broad open anticline is also the main trap geometry for the upper reservoir-seal system, but this is affected by a graben bounded by two oppositely dipping normal faults; one *sealed fault* and one *open fault*. An additional, subtle, low relief anticline forms an additional trap in the footwall of the graben-bounding sealed fault. The graben-bounding faults tips -out downdip into the basal layer of the upper reservoir (sand E) and updip into the base of the top regional seal (the uppermost sand layer in the model), as such they only affect the stratigraphy in the uppermost reservoir. The *sealed* and *open faults* have different properties and sealing potential: the *sealed fault* is designed as a sealing fault with a low permeability fault core, whereas the *open fault* has a high permeability fault core and would potentially act as a conduit for cross-formational fluid flow.

*Anticipated flow from well [9,3].* The buoyant gas phase flows upwards and reach the anticline sealing layer (sand ESF) above the injection point [9,3]. $CO_2$-saturated water is observed in the near-well region directly after onset of $CO_2$ injection. The anticline dipping angle facilitates gas migration into *Box A* and accumulation at the highest point of the CO₂ trap. The trap fills with gas and a layer with $CO_2$- saturated water forms underneath the downwards expanding gas accumulation. The $CO_2$-saturated water flows downwards into *Box C* over time due to i) the positive pressure gradient from the expanding gas and ii) convection because of the increased density relative to formation water. The gas accumulation increases upon continued injection until the gas-water interface aligns with the *spill point;* the excess gas flows through the *heterogeneous fault* and into *Box B* containing the fining upwards sequence and upper fault zone. The layered sequence (sands F, E, D and C, bottom to top) temporarily traps buoyant gas and laterally spreads the gas phase at the capillary barriers between layers. The increased density of $CO_2$-saturated water relative to the formation water leads to gravitational fingers. The $CO_2$ injection ends (after 305 min) when the gas reaches the upper sand layer (sand C) under the seal, and $CO_2$ in all forms is contained between the left no-flow boundary and the *sealed fault*.

*Anticipated flow from well [17,7].* The gas phase (injected in sand F) flows upwards and spreads laterally at layer boundaries in the fining upwards sequence (except between sand F and E, cf. **Table 2**). The gas phase advances upwards sequentially when it exceeds the capillary entry pressure in each layer. The $CO_2$-saturated water flows downwards due increased density and the pressure gradient of the gas accumulation – its flow pattern is influenced by the permeability variations in the layered sequence. The gas phase accumulates under the top seal above the injection well and migrates laterally until $CO_2$ injection is terminated (after 165 min). Depending on the amount of $CO_2$ injected, the gas phase will reach the *open fault*, and $CO_2$ in all forms will be contained between the *open fault* and the right no-flow boundary.

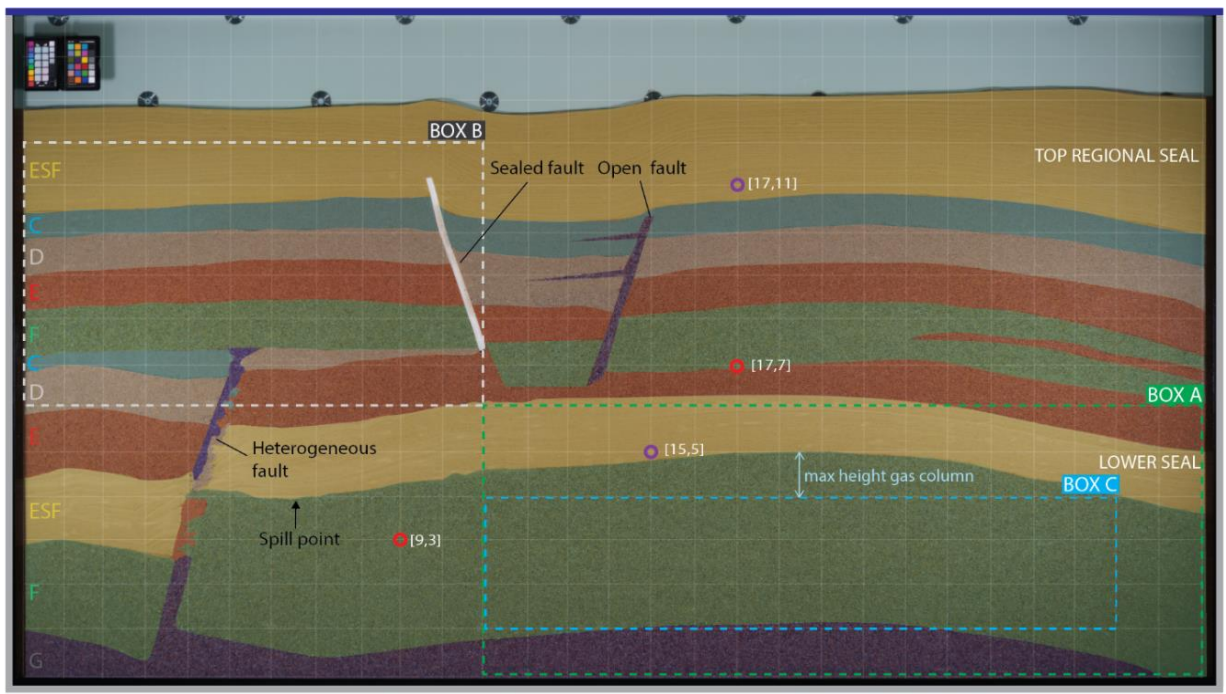

**Figure 1:** The benchmark geometry with color enhanced layers for facies identification. Each sand type (ESF, C, D, E, F and G; cf. **Table 2**) has a separate color indicated to the left. Sand/color correlation: ESF/yellow; C/light blue; D/light brown; E/red; F/green; G/dark blue. The geometry includes three faults: *sealed* (silicone strip), *open* (sand G), and *heterogeneous* (sands G, F, D and C). Total length of visible porous media is 2800 mm, and porous media height is nominally 1300 mm. Edge shadows visible on the left and right, and the active porous extends 30 mm behind the black metal frame on each side. The three no-flow boundaries (left, right and bottom) are indicated grey, whereas the open boundary is blue (top). A 100 x 100 mm cartesian grid with the origin [0,0] in the lower left corner with the x-axis positively oriented towards the right and the y-axis positively oriented towards the top aids the following coordination. Four monitored ports: two CO₂ injection well (red circles, coordinates [9,3] and [17,7]) and pressure ports (purple circles, coordinates [15,5] and [17,11]). Areas for reporting (Box A, B and C) are defined with the following coordinates (top right = TR; top left = TL; bottom right = BR; bottom left = BL): Box A:  TL [11,6] -> TR [28,6], BL [11,0] -> BR [28,0]; Box B: TL [0,12] ->TR [11, 12], BL [0,6] -> BR [11,6]; Box C: TL [11, 4] ->TR [26,4], BL [11,1] ->BR [26,1].

## 2.5 Image acquisition and analysis

The camera (Sony A7III, lens SAMYANG AF 45 mm F1.8) used the following settings (kept constant through all injections): shutter speed 1/30 sec; F number F2.8; ISO 100; color temperature 4100 K; and, manual focus). The camera was positioned in the curve focal point with a 3.6 m distance from the center point in the rig, halfway up the window height. Images were captured at high spatial (7952 x 4472 pixels, for a total of 35.5 megapixels) and temporal (between 10 sec to 5 min intervals, depending on active experimental phase) resolution to capture displacement and mass-transfer dynamics. Each run consists of more than 1000 images; a subset that captures key events, displacement processes and mass-transfer dynamics is available for open-access download (Eikehaug et al 2023b). The subset contains 137 high-resolution images with the following intervals: 10 images before $CO_2$ injection at 20 second intervals; images every 5 min during the first 360 min (6 hours) of the experiment (73 images); images every hour until 48 hours (42 images); images every 6 hours until end of experiment (12 images).

*Image analysis toolbox*

To use the high-resolution images as measurement data, image analysis is required. As part of the benchmark study, the open-source image analysis software *DarSIA* (short for <u>Dar</u>cy <u>S</u>cale <u>I</u>mage <u>A</u>nalysis, Both et al, 2023a) has been developed, detailed in [Nordbotten et al 2023, this issue]. DarSIA provides the capability to extract physically interpretable data from images for quantitative analysis of the image sequences of the time-lapsed $CO_2$ injection and storage experiments. In particular, DarSIA includes preprocessing tools to align images; project suitable regions of interest of images onto two-dimensional Cartesian coordinate systems; correcting for geometrical discrepancies due to e.g. the curved nature of the physical asset; as well as correcting white balance fluctuations and perform color correction utilizing the color checker attached to the physical asset, overall, resulting in unified image sequences. Furthermore, additional analysis tools are available to e.g. determine spatial deformation maps comparing different configurations and extract concentration profiles or identify phases, to mention a few. The latter aims at a Darcy scale interpretation of the high-resolution images taken of the physical asset, effectively, removing sand grains and upscaling fluid quantities.

*Phase identification*

The image analysis toolbox was used to separate between the different $CO_2$ phases (gaseous and aqueous) present in the experiments, and a set of assumptions enabled the quantification of each phase to be calculated during the $CO_2$ injection and associated mixing. Four main phases are anticipated:

1. Free gas (potentially flowing gas phase with non-zero gas permeability, referred to as mobile gas)
2. Trapped gas (residually trapped $CO_2$ with zero gas permeability, referred to as immobile gas)
3. $CO_2$-saturated water (aqueous phase with a non-zero $CO_2$ content)
4. Formation water (aqueous solution with zero $CO_2$ content)

A range of assumptions (cf. **SM 3**) was needed to quantitatively describe the observed multiphase flow phenomena during repeated $CO_2$ injections in the physical flow rig. Based on these assumptions, a geometric separation of the formation water from any $CO_2$ in the system and of the gaseous $CO_2$ from the $CO_2$-saturated water is sufficient. This separation was possible due to the use of the pH-indicator mix (cf. **Table 1** and **Figure 2**). Through pixel-wise image comparison to the image corresponding to the injection start, a thresholding approach both in terms of monochromatic color space and signal intensity accomplishes the separation (in addition further techniques are used to convert the signal to Darcy-scale quantities, cf. Nordbotten et al 2023, this issue). The heterogeneous nature of the geometry is considered in the analysis by choosing facies-based threshold parameters, and thereby allows for tailored and relatively accurate phase segmentation, cf. Figure 2. The parameters are chosen such that transition zones are included as demonstrated. The same unified setup has been used for analyzing all experimental runs.

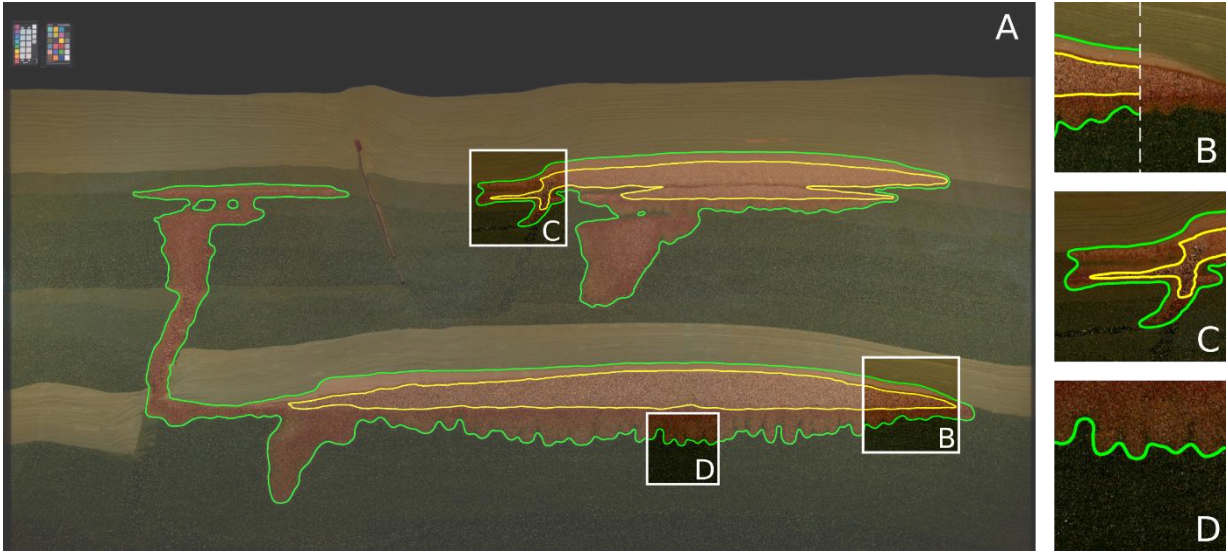

**Figure 2.** Resulting phase identification of formation water, $CO_2$-saturated water and free gas using DarSIA, at injection stop; two plumes are identified, containing free gas regions (yellow contour) and $CO_2$-saturated water (green contour). Subfigure B: The pH-indicator mix (left and right, with and without contours, resp.) allows for visual separation of the different phases based on color spectra. Subfigure C: Detection of free gas in the open fault. Subfigure D: Due to the use of regularization in the upscaling, DarSIA smears out fingers and thus merely detects fingertips for fingers that are closer than a few grain diameters.

It must be noted, that based on the choice of the assumptions and the resulting image analysis, the identification of gaseous phases for which assumption **SM 3.I** is not satisfied, may be erroneous; transition zones smear out and the saturation decays which leads to a sudden disappearance of the post-processed gaseous phase due to the use of fixed threshold parameters. In all experimental runs, two gaseous regions are detected, cf. **Figure 2**, and the described effect takes place for the upper gaseous region, whereas the lower region is detected stably. While the upper region fully dissolves, the lower region results in remaining gas, cf. **SM 3.III**, which is detected as gaseous $CO_2$. Consequently, the subsequent quantitative analysis reports on a small amount of non-vanishing gas accumulation towards the end of the experimental runs.

*Procedure in the quantitative analysis*
The subsequent quantitative analysis results from post-processing the phase identification. We briefly elaborate on the procedure of key computations.

1. *Mass calculations and concentration maps.* Total $CO_2$ mass of dissolved and mobile $CO_2$ are determined through integration of the pixel-wise defined areal densities of mobile $CO_2$, $m_{CO_2}^g = \phi \cdot d \cdot s_g \cdot \chi_c^g$, and dissolved $CO_2$, $m_{CO_2}^w = \phi \cdot d \cdot s_w \cdot \chi_c^w$, with the single components determined as follows. Based on assumption **SM 3.V**, the porosity $\phi$ and the depth $d$ can be accurately determined. Resulting from assumption **SM 3.I**, the phase identification provides saturation maps $s_g$ for the gaseous phase and $s_w$ for the aqueous phase, taking values either 0 or 1. It remains to quantify the mass concentrations of $CO_2$, $\chi_c^g$ and $\chi_c^w$ in gaseous and aqueous phases, respectively. Based on assumption **SM 3.I**, $\chi_c^g$ is provided as the density of gaseous $CO_2$ under operational conditions, cf. **SM 1**, obtained from the NIST database (Lemmon et al. 2022). With that, the pixel-wise areal density $m_{CO_2}^g$ is known. Assumption **SM 3.II** allows now for obtaining the remaining mass concentration $\chi_c^w$ through sparsification, as follows. As illustrated in **Figure 2**, two $CO_2$ plumes originating from the two injection ports remain unconnected throughout almost the entire run time (until 84 hours). The total $CO_2$ mass in each plume is known at any point in time based on the injection protocol, cf. **SM 4**, while the respective total mass of mobile $CO_2$ is determined through integration of $m_{CO_2}^g$ over the area of the plumes. Subtraction of both provides the total mass of dissolved $CO_2$ for each plume. Finally, by assumption **SM 3.II**, $\chi_c^w$ set to be 0 in the formation water; constant and equal to the proportionality constant between the total volume and the total mass of dissolved CO₂ in each connected region of CO₂-saturated water $\chi_c^w$; and not relevant for the mass calculations, yet for the discussion of convective mixing, in the remaining gaseous regions, $\chi_c^w$ is set to $\chi_{c,max}^w$ =1.8 kg/m³.
2. *Physical variability.* Given a set of phase segmentations, associated to different configurations, the intersections and complements of phase segmentations can be directly determined. Furthermore, we introduce metrics based on volume-weighted ratios of these, to quantify corresponding overlap and unique appearances of detected regions.
3. *Fronts and fingers.* When restricted to a region of interest, the internal interface between the detected water formation and the CO₂-saturated water can be interpreted as propagating front. Its length can be determined by making use of the Cartesian coordinate system attached to the images. Extremal points can be identified as fingertips, allowing to count them over time. Due to the use of regularization in DarSIA, when converting grain-scale data to Darcy scale, fingers are slightly smeared out. This affects the detection of the free space in between fingers, cf. **Figure 2**. Thus, in these regions the resulting interface between the formation water and CO₂-saturated water can be understood as approximating non-convex hull of the fingers with its length being a lower estimate to the actual contour length of the fingers. The detection of single fingertips is however not affected resulting in lower uncertainty.

## 3. Results and Discussion

This section is divided into two parts: Part 1 relates to the sparse dataset requested in the benchmark study [Flemish et al 2023, this issue], and includes a discussion on temporal behavior for studied parameters across repeated runs; Part 2 expands our analysis and focuses on physical variability between repeated injections, drivers for the observed variability.

### 3.1 The benchmark sparse data set

The sparse data set [Nordbotten *et al.* 2022] requested six data points to assess the ability of the participating modeling groups to forecast relevant properties of the physical system. The $CO_2$ phase was to be reported in the following three categories: *mobile free phase* (gas at saturations with a positive gas relative permeability), *immobile free phase* (gas at saturations with zero gas relative permeability), *dissolved* (mass of $CO_2$ in $CO_2$-saturated water). The sum of the *mobile*, *immobile* and *dissolved* phases equals the total mass of $CO_2$. The sparse data set is included for completeness here, but the reader is referred to [Flemish et al 2023, this issue] for comprehensive analysis and discussion.

The following sparse data were requested (cf. **Figure 1** for described regions and pressure ports):

1. *As a proxy for assessing risk of mechanical disturbance of the overburden*: Maximum pressure [N/m$^2$] at pressure port a) [15,5] and b) [17,11].
2. *As a proxy for when leakage risk starts declining*: Time [s] of maximum mobile $CO_2$ [g] in Box A.
3. *As a proxy for our ability to accurately forecast near-well phase partitioning*: $CO_2$ mass [g] of a) mobile; b) immobile; c) dissolved; and d) total in seal in Box A at 72 hours after injection start.
4. *As a proxy for our ability to handle uncertain geological features*: $CO_2$ mass [g] of a) mobile; b) immobile; c) dissolved; and d) total in seal in Box B at 72 hours after injection start.
5. *As a proxy for our ability to capture onset of convective mixing*: Time [s] for which the quantity

$$M(t) \equiv \int_C \left| \nabla \left( \frac{\chi_c^w(t)}{\chi_{c,\max}^w} \right) \right| dx$$

   first exceeds 110% of the width of Box C, where $\chi_c^w$ is the mass fraction of $CO_2$ in $CO_2$-saturation water.
6. *As a proxy for our ability to capture migration into low-permeable seals*: Total mass of $CO_2$ [g] in the top seal facies (sand ESF) at final time within Box A.

Here we report laboratory sparse dataset (cf. **Table 3**) using the dataset [Eikehaug et al. 2023b] and dedicated DarSIA scripts (Both et al 2023b) with assumptions (cf. **SM 3**). The $CO_2$ distribution after 72 hours with locations of Box A, Box B and Box C is included to aid interpretation (see **Figure 3**).

Table 3. Benchmark sparse dataset

| Parameter | 1. | | 2. | 3 [a]. | | | | 4 [a]. | 5. | | 6. |
|---|---|---|---|---|---|---|---|---|---|---|---|
| Units | 10$^5$ [N/m$^2$] | | 10$^3$ [s] | 10$^{-2}$ [g] | | | | 10$^{-2}$ [g] | 10$^3$ [s] | | 10$^{-2}$ [g] |
| Sub Parameter | 1a | 1b | 2 | 3a upper | 3a lower | 3c | 3d | 4c | 5 upper | 5 lower | 6 |
| C1 | 1.11 | 1.05 | 15.4 | 16.3 | 2.9 | 320 | 37.6 | 76.3 | 12.9 | 12.2 | 54.3 |
| C2 | 1.12 | 1.06 | 14.7 | 27.2 | 10.1 | 307 | 36.7 | 76.0 | 14.1 | 12.4 | 62.2 |
| C3 | 1.14 | 1.08 | 14.9 | 28.1 | 12.5 | 313 | 40.4 | 71.5 | 14.9 | 12.8 | 59.1 |
| C4 | 1.11 | 1.05 | 15.1 | 18.9 | 5.0 | 313 | 38.5 | 74.6 | 15.4 | 13.3 | 55.4 |
| C5 | 1.11 | 1.05 | 13.7 | 25.3 | 6.9 | 298 | 37.8 | 90.5 | 17.3 | 12.3 | 52.6 |
| Average ± 1 std | 1.11 ± 0.01 | 1.05 ± 0.01 | 14.8 ± 0.6 | 23.2 ± 4.7 | 7.5 ± 3.5 | 310 ± 7 | 38.2 ± 1.2 | 77.8 ± 6.6 | 14.9 ± 1.5 | 12.6 ± 0.4 | 56.7 ± 3.5 |
| Uncertainty | ± 10$^{-3}$ | ± 10$^{-3}$ | ± 0.06 | - | | ±20%[b] | ± 20 %[b] | ± 20 %[b] | - | | ± 20 %[b] |

[a] Parameters 3b, 4a, 4b and 4d are reported as zero and not included in Table 3. Rationale provided in main text below.
[b] Stated uncertainties are discussed in the main text below.

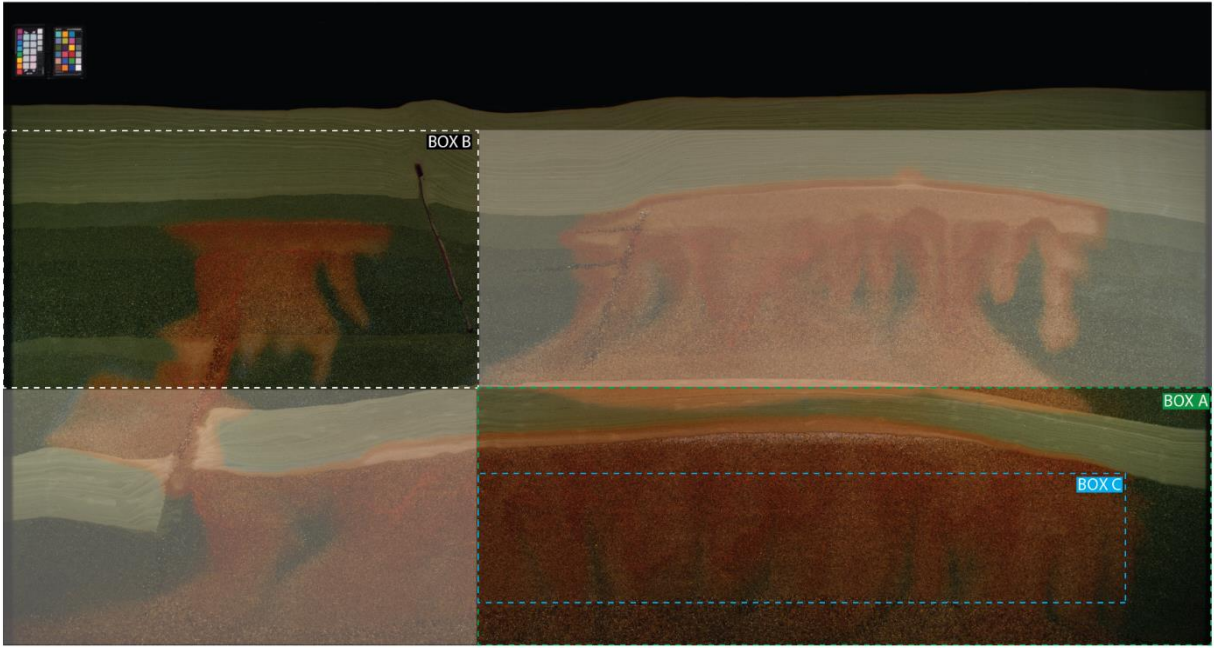

**Figure 3.** Distribution of $CO_2$ after 72 hours for run C3. The positions of Box A (green, dashed line), Box B (white, dashed line) and Box C (blue, dashed line) are used to populate the sparse benchmark data set. The shaded regions in the benchmark geometry (top right and bottom left) are outside the defined boxes. $CO_2$ (in any form) in the shaded regions was not included in the analysis for the sparse data set.

*Maximum pressure at ports [15,5] and [17,11] (parameters 1a and 1b).* The maximum pressures at the pressure ports ([15,5] and [17,11]) located in the sealing structures (sand ESF, cf. **Figure 1**) were initially recorded with five pressure transducers (ESI, GSD4200-USB, -1 to 2 bara) because single digits millibar pressure gauges were not available for the benchmark study. The results were, however, discarded because 75% of the transducers recorded pressures less than the atmospheric pressure in the room. Hence, we use historical atmospheric pressure data reported from a nearby meteorological weather station (cf. **SM 1**), and adjust for differences in height and hydrostatic pressures (see **Table 3**). We apply an uncertainty of ± 1 mbar, five times stated instrument accuracy, to account for the possible overpressure during $CO_2$ injections.

*Time of maximum mobile $CO_2$ in Box A (parameter 2).* The development in mobile gas in Box A for all five runs (cf. **Figure 4**) increased linearly with the injection rate until the gas accumulation aligned with the spill point (defined in **Figure 1**). On average, the maximum mass of mobile gas was observed after 4.11 ± 0.17 hours. While there appears to be is some noise in the identification of the mobile gas, the time of maximum value is a clearly defined peak in the time series. Seen together with temporal resolution of the image series (20 seconds per frame), we expect the uncertainty of our identification of the time of maximum mobile $CO_2$ to have an uncertainty of no more than three frames, i.e. ± 1 min. The nature behind the fluctuating mass after the initial spill (cf. black rectangle, **Figure 4**) is discussed in more detail in **chapter 3.2**.

*Mobile, immobile and dissolved $CO_2$ in Box A and Box B (parameters 3, 4 and 6).* The mass of mobile gas in Box A (parameter *3a* in **Table 3**) was on average 0.232 ± 0.047 g, and is considered an upper bound for this parameter. The lower bound was found indirectly from the observation of non-zero mass of mobile gas at the end of the experiments (cf. **Figure 4**), related to atmospheric gases in the formation water due to insufficient degassing (cf. **chapter 2.1** and Haugen et al 2023, this issue). Based on our physical understanding of the studied system, we anticipate that the mass of mobile $CO_2$ should be zero at the end of the experiment. Hence, we subtract the end point mass from the upper bound to find an estimate of the lower bound, cf. **Table 3**. An alternative, but also physically plausible, lower bound for parameter 3a is zero, where all the mobile gas ($CO_2$) is dissolved in the $CO_2$-saturated water. The mass of mobile gas in Box B after 72 hours (parameter 4a) is reported as zero because mobile gas was not observed in the segmented images.

The mass of immobile gas in Box A and Box B (parameters *3b* and *4b* in **Table 3**) were reported as zero because the formation water did not generate a unique and characteristic color for immobile gas. Hence, DarSIA and its color/signal-based segmentation (cf. chapter 2.5) is not able to distinguish immobile gas from the other phases. Careful visual inspection identified small amounts of immobile gas at early times, but visual inspection at 72 hours did not identify any immobile gas. This is consistent with our physical understanding of the system, where isolated gas bubbles are expected to dissolve quickly.

The mass of dissolved gas in the $CO_2$-saturated water in Box A and Box B after 72 hours (parameters *3c* and *4c* in **Table 3**) were 3.10 ± 0.07 g (Box A) and 0.778 ± 0.066 g (Box B), see **Figure 5**. The mass calculations use the known injected $CO_2$ mass in well [9,3] for Box A and well [17,11] for Box B, and apply DarSIA to segment the separate plumes originating from each well to calculate the mass of mobile and dissolved gas (cf. chapter 2.5). The two plumes remain unconnected throughout almost the entire run time (until 84 hours), and the total $CO_2$ mass in each plume is known at any point in time based on the injection protocol. After 84 hours the plumes merge and the plots are extrapolated to 120 hours (end of experiment) based on current trends. The mass of $CO_2$ in the sealing structures in Box A and Box B after 72 hours (parameters *3d* and *4d* in Table 3) were 0.382 ± 0.012 (Box A, cf. **Figure 5**) and 0.00 (Box B). Mobile and dissolved gas did not enter the top regional seal confined within Box B, but minute amounts of dissolved gas (in the order of $10^{-3}$ g) entered the sealing structure in the lower, right corner of Box B after 72 hours. Hence, the final mass of $CO_2$ in the sealing structure confined within Box A (parameter *6*, cf. **Figure 6**) was on average 0.567 ± 0.035. For the parameters discussed here (3c, 3d, 4c, 4d and 6) we attribute a nominal measurement uncertainty of ± 20 % based on the limitations and influence of underlying assumptions (cf. **SM 3**), stated weakness in the analysis of the color scheme (cf. chapter 2.5), extrapolating trends and operational difficulties with mineralization of methylene red.

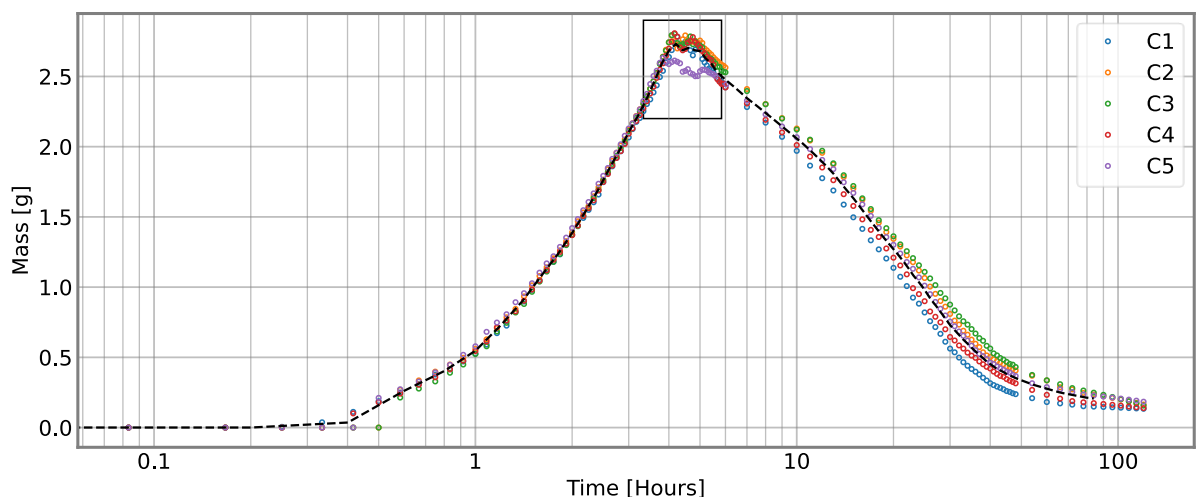

**Figure 4**. Development in mass [g] of mobile gas in Box A for the whole experimental time (120 hours) for all five runs (C1 – C5) and the average (black, dashed line). The mass increased linearly with the injection rate until spill time (cf. **Table 3**), and then decreased because the mobile gas dissolved into the formation water. The development in mobile mass associated with the spill point (black rectangular) is discussed in detail below.

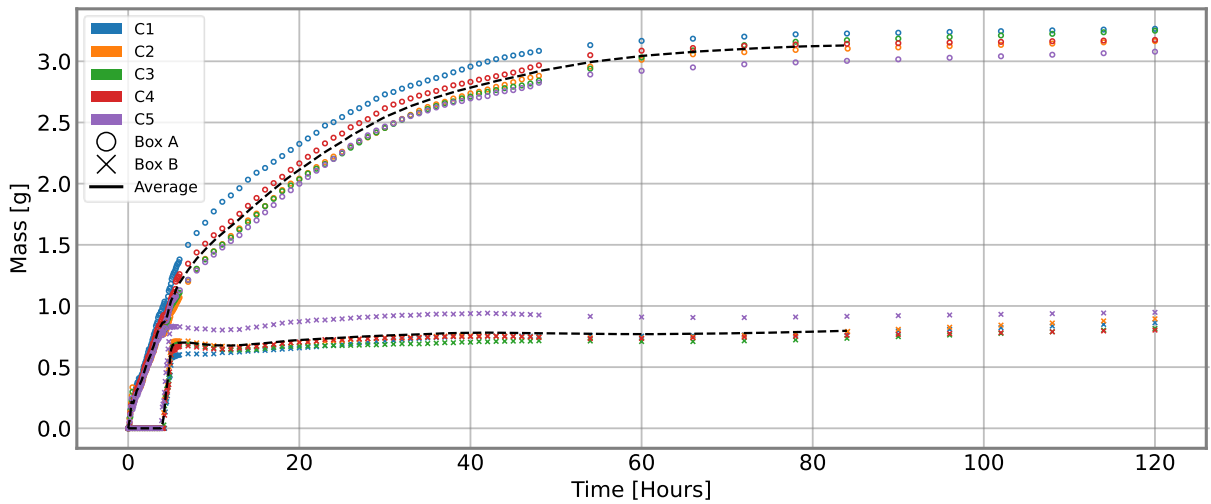

**Figure 5.** The development in mass of dissolved $CO_2$ [g] in $CO_2$-saturated water in Box A (open circles) and Box B (crosses) for runs C1-C5 during the whole experimental time (120 hours). All mass curves increase from the onset because mobile gas dissolved into the formation water to form $CO_2$-saturated water and reach plateau values when most of the gas within each box is dissolved. The curves in Box B remain zero until the gas exceeds the spill point and flow into the fault (after approximately 4 hours). The somewhat different development for run C1 in Box A (blue circles) and run C5 in Box B (purple crosses) relates to the inconsistencies for these runs, discussed in chapter 3.2. Note that the average curves (black, dashed lines) are calculated until 84 hours.

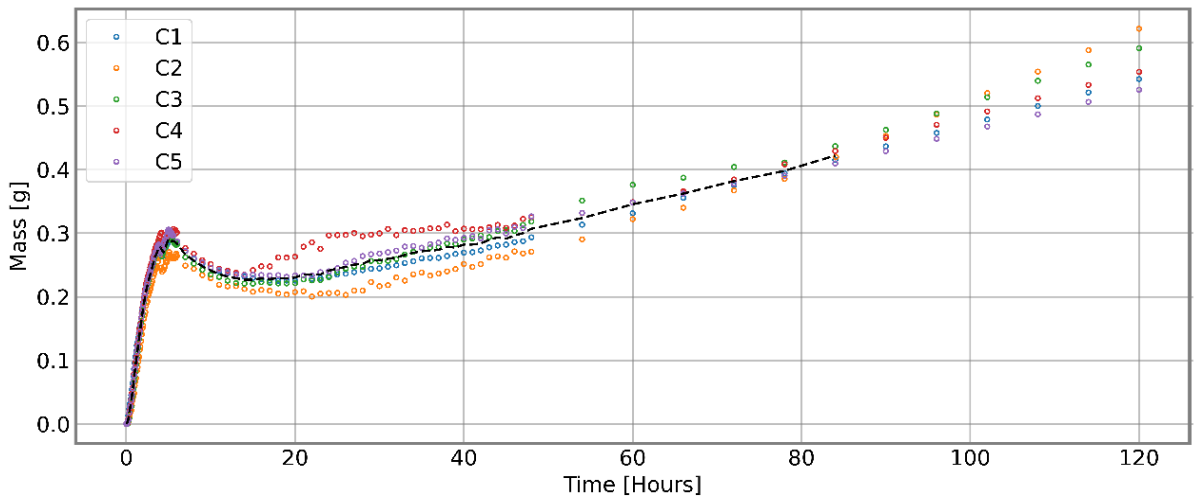

**Figure 6.** Development of $CO_2$ (in any form) in sealing layer (sand ESF) confined within Box A during the whole experimental time (120 hours) for all five runs (C1-C5). Only $CO_2$-saturated water (no gas) was observed in the sealing layer in Box A, and advection from the underlaying gas was the main driving force for increased mass initially. After gas injection stopped (after approximately 5 hours), there was a slight decrease of $CO_2$ mass in the sealing layers, explained by gravity of the denser $CO_2$-saturated water and diminishing advective forces due to a reducing gas cap under the anticline. After approximately 20 hours, the mass increase again because $CO_2$-saturated water from injector [17,7] flows downwards and enters the top boundary of Box A (cf. **Figure 4** after 72 hours).

*Development in M (t) relative to the width of Box C (parameter 5).*

The $M(t)$ (parameter 5 in Table 3) is a measure of the total variation of the concentration field. As such, it is related to the contour lengths of the density driven fingers, and we normalize it relative to the length of Box C, so that a value of $M_{norm}(t) = 1$ corresponds to no fingers below a gas cap spanning the whole length of the top of Box C. As $CO_2$-saturated water migrated downwards due to gravity, the contour lines and the $M_{norm}(t)$ increase (see **Figure 7**). On average for the five runs $M_{norm}(t)$ exceeds 110% of the width of Box C after 4.14 ± 0.4 hours, where the stated times for each run may be considered as an upper bound due to the assumption that the concentrations are constant, which decreases the measure of the gradient in the integral. A lower bound is the time when $M_{norm}(t)$ reached 100% of the length of Box C, which is closely correlated to gas filling the upper boundary of Box C, a necessary prerequisite for $M_{norm}(t)$ exceeding 110%.

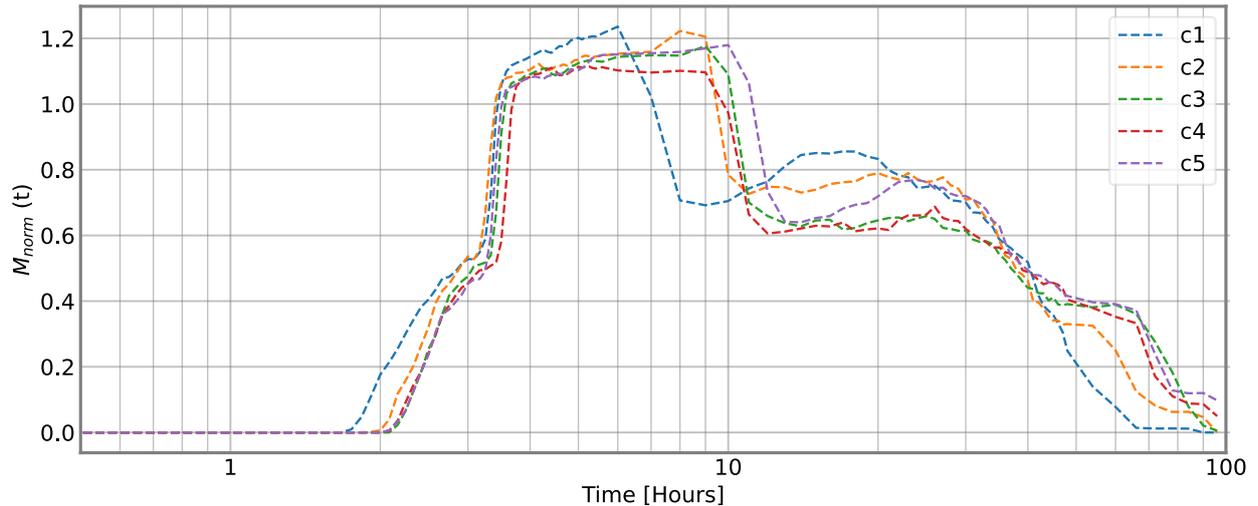

**Figure 7.** Development in $M_{norm}(t)$ for all five runs from injection start until end of experiment (120 hours). For the initial state of a zero CO₂ concentration within Box C, $M_{norm}(t)$ takes the value 0. Run C1 (blue) is ahead of the other runs, both in the start and at the end (fingers start to leave Box C). The rapid increase between 3 and 4 hours arises because the mobile gas fills the top of Box C. The reverse is true after approximately 10 hours (6 hours for run C1) when the gas accumulation (due to shrinking by dissolution) exits the upper boundary of Box C and the parameter $M_{norm}(t)$ rapidly decreases. This is counterbalanced to some extent by the further development of the density-driven fingers, as seen around 20 hours, until dissolution and diffusion eventually leads to a more uniform distribution of dissolved CO₂, and $M_{norm}(t)$ approaches 0 again.

**3.2 Physical repeatability of multiphase flow during laboratory carbon sequestration runs**

The benchmark study consisted of five operationally identical $CO_2$ injection experiments using the same geological geometry and initial conditions. The experiments were designed to generate physical data for model comparison, with the motivation to achieve a physical 'ground truth'. Here we discuss physical repeatability between the five runs (C1-C5) by comparing the degree of areal sweep overlap incorporating all forms of $CO_2$ (mobile, immobile, dissolved) in three regions (Box A, Box B' and Box D, cf. **Figure 8**) with increasing geological complexity. We quantify the degree of overlap of runs C2, C3 and C4, and discuss the uniqueness of each run.

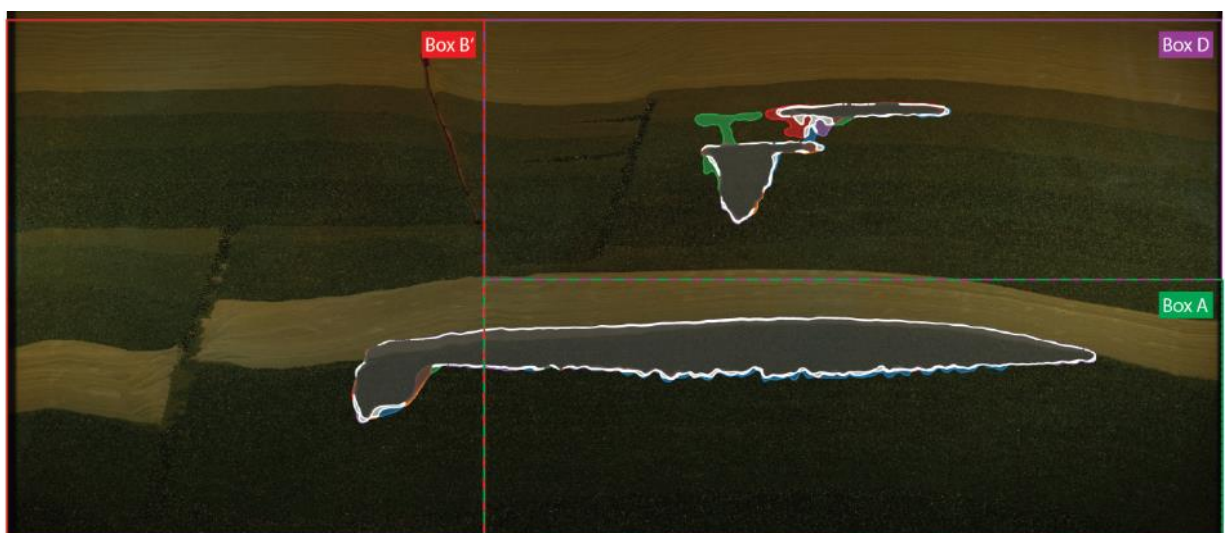

**Figure 8**. Degree of physical overlap and description of Box A, Box B' and Box D with increasing geological complexity. Box A is identical to Figure 1; Box B' is an extension of Box B (cf. **Figure 1**) and includes the lower part of the geometry left to the heterogenous fault; Box D includes the fining upwards sequence associated with injector [17,7] and the open fault (cf. **Figure 1**). The CO₂ distribution (all forms) for all five runs (C1-C5) in three

boxes (Box A, Box B' and Box D) after 155 min of $CO_2$ injection. Spatially distributed overlap for all runs, with the following color scheme: gray (overlap C2+C3+C4); blue (unique C1); orange (unique C2); green (unique C3); red (unique C4); purple (C5 unique); brown (combinations all runs with at least one of C2, C3 or C4), white (other combinations). The reader is referred to **SM 5** for additional time steps.

Physical reproducibility with increasing reservoir complexity

We investigate the reproducibility between five runs in the same geometry, with the hypothesis that increased reservoir complexity tends to reduce the degree of physical reproducibility. As mentioned above, our motivation to achieve a physical "ground truth" was not fully achieved. This was because our 'identical' experiments indeed were not truly identical, even if the gas injection protocol was (within measurement uncertainty, cf. **SM 4**). Next, we describe the two known variables that influence the displacement patterns:

1. *Inconsistent water chemistry*. The formation water (cf. **Table 1**) in run C1 unintentionally used tap water instead of deionized water. The inconsistent water chemistry for C1 resulted in a unique dissolution rate and convective mixing behavior (cf. **Figure SM.3**). Run C1 is thus omitted from the analysis of physical reproducibility.
2. *Atmospheric pressure variations.* The atmospheric pressure variations in Bergen (cf. **Figure SM.1)** resulted in a low-pressure outlier for run C5 (968 mbar) compared with the other runs (on average 999 mbar injection period, cf. **Table SM 1**). Hence, the larger volume of the injected $CO_2$ (equal mass injected for all runs) influenced key parameters in the experiment (most prominently parameter 2 in **Table 3**, but also rate of dissolution). Run C5 is thus omitted from the analysis of physical reproducibility.

The described operational (water chemistry) and environmental (atmospheric pressure) inconsistencies provide the rationale for excluding C1 and C5 in our analysis of physical reproducibility for operationally identical experiments with comparable pressure and temperature conditions. An analysis of sand settling between runs showed only minor changes (cf. **SM 6**). Hence, we focus on runs with comparable system parameters, and report the development in overlap between runs C2, C3 and C4 (cf. **Figure 9**). To compute the overlap percentages, we first weight all pixels in the segmented images with their corresponding volume (see **SM 2**). Then, the ratio between the number of volume-weighted pixels where $CO_2$ (gas and dissolved) in C2, C3, and C4 overlap and the number of volume-weighted pixels where $CO_2$ (gas and dissolved) in any of the three runs appear is reported. Next, we describe the development in physical overlap within Box A, Box B' and Box D.

The development in physical overlap in Box A may be divided into four intervals: *i. pre-spilling*; *ii. gravitational fingers*, *iii*. *dissolution-driven flow* and *iv*. *homogenization.* The *pre-spilling* interval (from the injection start to approximately 4 hours) occurred before the gas column height exceeded the spill point. The onset of gravitational fingers occurred in this interval, but they are still only minor and do not develop into pronounced gravitational fingers. The overlap increased from injection start and reached a global maximum (97 % overlap) after approximately 4 hours, with an average 92 % $C_{2,3,4}$ overlap for the whole interval. The uniqueness of runs C2, C3 and C4 were on average 0.14 % (cf. **Figure SM.4**) during the *pre-spilling* period. The *gravitational fingers* interval (approximately 4 to 30 hours) was characterized by development of pronounced gravitational fingers under the gas accumulation in the anticline trap in Box A. The physical overlap of $C_{2,3,4}$ decreased from 97 to 79 % (local minimum), dominated by the differences in number of fingers and individual finger dynamics (discussed in more detail below). The *dissolution-driven flow* interval (approximately 30 to 70 hours) describes the period when the gravitational fingers reached the no-flow at the lower Box A boundary, and fingers start to move lateral and merge as the gas accumulation dissolves and pull aqueous phase from surrounding regions into Box A. The physical overlap increased to above 95 % in this period. The *homogenization* interval (approximately 70 to 120 hours) was characterized by a constant physical overlap (above 95 %) with only minor movement of aqueous phases confined in Box A.

Box B' generally follows the overall behavior of Box A in the four intervals defined above. Importantly, the reduction in physical overlap observed in the *gravitational fingers* interval (after approximately 4

hours) was related to variable spilling times for runs C2, C3 and C4, not related to finger development (cf. parameter 2, **Table 3** that approximates the spilling time for each run). The variation in spill times resulted initially in reduced overlap with slight variation in fault migration and displacement patterns for runs C2,C3 and C5. The sustained reduction of physical overlap stems from an apparent stochastic variation for run C3 (cf. **Figure SM.3**; 10 hours), corroborated with development of the uniqueness for each run (cf. **Figure SM.4**; middle). The physical explanation for the observed variation in run C3 is not clear, but this only occurred for that single run, with subsequent runs (C4 and C5) reverting to the flow patterns seen for the earlier runs (C1 and C2). Hence, we do not expect the deviation in run C3 to stem from any physical alterations within the experiment (sand settling, or chemical alterations). Remaining explanations could be related to variations in atmospheric pressure, or factors outside our experimental control.

The development in Box D was delayed in time relative to Box A and Box D due to the later injection start of well [17,11], but follows the overall trend: initially increasing overlap, slight reduction due to finger development and convective mixing, then increase through homogenization. Small amounts of dissolved gas were observed in a localized point the top regional seal contained in Box D for most runs (cf. **Figure SM 3**). The seal breach occurred around a plugged port ($CO_2$ migrated along the sealing silicone), resembling a of CO2 leakage scenario along a poorly abandoned well.

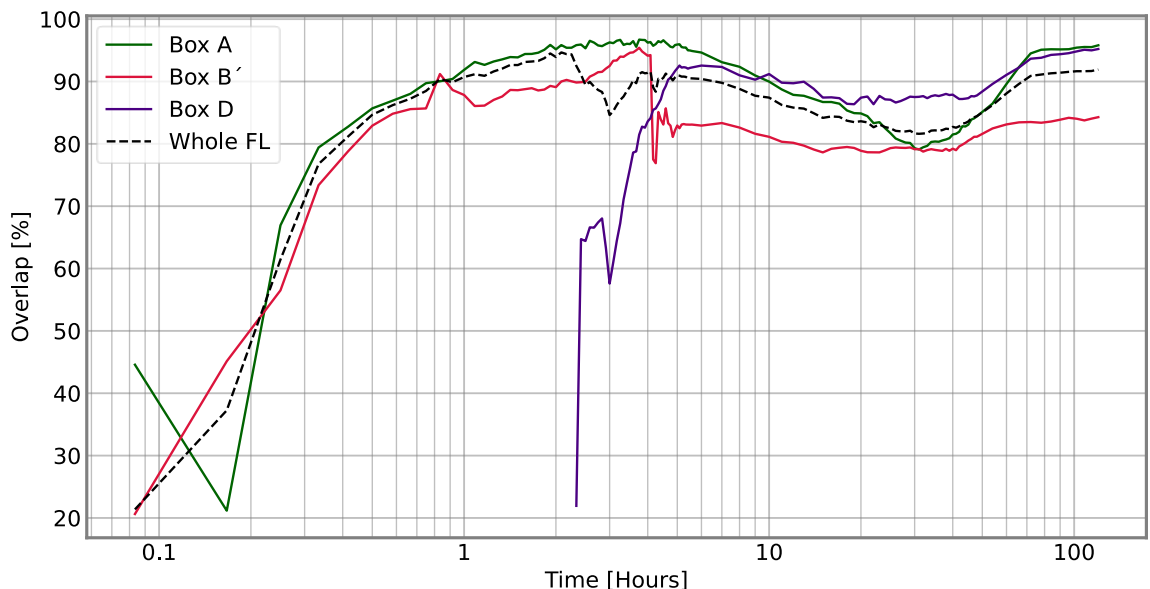

**Figure 9.** Degree of physical reproducibility between operationally identical $CO_2$ injection runs with comparable pressure and temperature conditions (runs C2, C3 and C4). Box A (green line) represents the most homogenous case; Box B' (red line) represents the case with the heterogenous fault zone and fining upwards sequence; Box D (purple line) represents the middle case with a fining upwards sequence. Overlap considering the whole geometry (dashed line) is included for comparison.

Dynamics of gravitational fingers in Box C
Box C is the homogenous zone under the lower anticline under the main gas accumulation and where most of the gravitational fingers emerge during and after $CO_2$ injection. From image analysis it was possible to extract the development of fingers as a function of time for all runs (cf. **Figure 10**). The fingers appear after an onset time of approximately 3 hours, and the number is reasonably stable around 25-30, which corresponds to a characteristic spacing of about 5-6 cm. The stability of the number of fingers is an indication that the system is near the regime of the "maximally unstable" fingers spacing, predicted by theoretical considerations (see e.g. Riaz et al 2006; Elenius et al 2012). This observation is supported by the finger lengths, which indicate a linear growth regime after onset. Repeatability was observed in terms of onset location and finger dynamics, even at time significantly after onset (cf. **Figure SM.5** and **Table SM.3**).

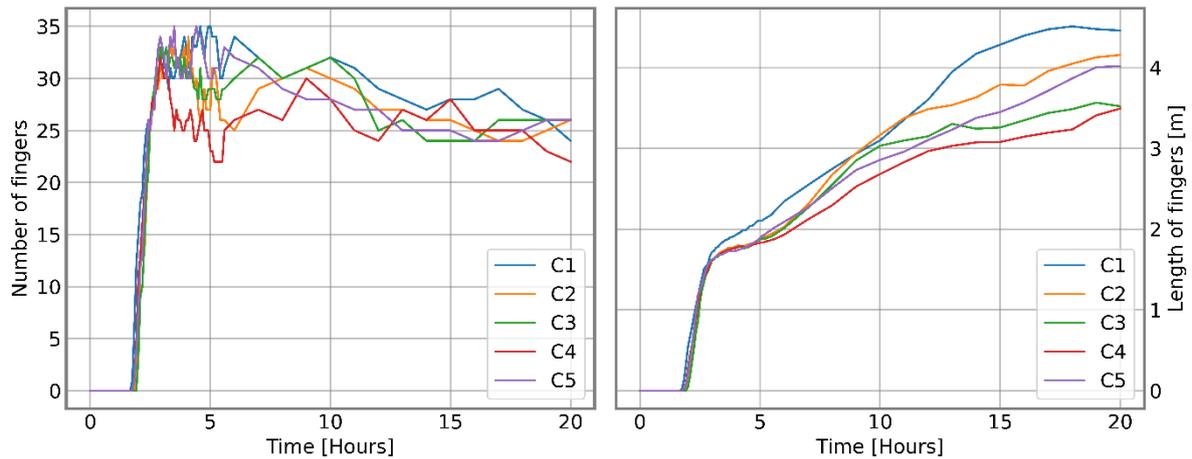

**Figure 10.** Dynamics of convective mixing and gravitational fingers in Box C for all runs C1-C5. <u>Left</u>: Number of gravitational fingers, all runs follow the general trend: a rapid increase until a maximum is reached, followed by a declining number as some fingers merge. <u>Right</u>: The length [m] of the boundary of the phase segmentation, also identifying (an approximation) of the fingers. Note that the contour length only considers the boundary inside Box C. Both graphs end when the first finger reached the lower boundary of Box C (20 hours).

Oscillating $CO_2$ leakage from anticline

The benchmark geometry and injection protocol were designed to achieve realistic displacement processes relevant for subsurface carbon storage, where most observed phenomena and mass transfer dynamics were anticipated; showcased in the description of expected behavior (cf. **chapter 2.4**) and benchmark description [Nordbotten et al 2022]. An oscillating $CO_2$ spilling event from the lower anticline was observed in our study, something that was not anticipated. Non-monotonic leakage behavior has previously been suggested in the literature (Preuss 2005), and in natural analogues (Shipton et al, 2004), attributed to the interplay between multiphase flow, Joule-Thomson cooling, and heat transfer effects in the fault plane. To our knowledge, oscillating $CO_2$ leakage behavior from an anticline into a fault zone in the absence of thermal effects has not previously been observed experimentally nor received attention in the literature. Below we discuss the displacement dynamics during multiphase flow in the fault plane generating the observed oscillating anticline $CO_2$ leakage behavior.

The mass of mobile gas in Box A oscillated after the initial spilling event for all runs (cf. **Figure 11**). The gas escapes the anticline trap in bursts and flows into the narrow restriction at the bottom of the fault (aligned in height with the spill point). When gas migrates upward in the fault zone (essentially a localized permeable pathway), it displaces resident aqueous fluids downwards. The inflow of aqueous phase effectively reduces and ultimately blocks the upwards migration of gas. This is in essence because the localized pathway in the inlet region of the fault cannot accommodate stable counter-current flow (upwards gas flow and downwards water flow), possibly due to viscous coupling effects (see e.g. the review paper by Ayub and Bentsen, 1999). When the upwards migration of gas is temporarily blocked, the anticline gas column height increases again with continued $CO_2$ injection. The process then repeats itself when the aqueous phase flow dissipates. A secondary effect is that the inflowing aqueous phase increases the local water saturation between the spill point and the inlet point of the fault and traps gas. The gas quickly dissolves into aqueous phase, and the subsequent spilling events (up to four events per run) are essentially local drainage processes, characterized by oscillating mass of mobile gas under the anticline (Box A). Interestingly, the process appears hysteretic in nature, with decreasing peak mass values for each event, most likely related either to increased gas relative permeability between the spill point and the fault, or changes in the local $CO_2$ concentration in the aqueous phase. The fluctuations stopped when the $CO_2$ injection terminated (after approximately 300 min, cf. **SM 4**), and the gas column height (and, hence, the mass of mobile gas) decreased under the spilling point.

To generalize the underlying causes for the observed phenomenon is difficult based on the reported experiments alone, and should be coupled with dedicated numerical simulations including more effects. The observations are to some degree influenced by the physical system (no-flow boundaries in the vicinity of the spill point and fault, and the fault geometry aligned with the spill point acting as a restriction of upwards migration of gas) and presence and shape of the gas accumulation effectively reducing the area available for water flow. A systematic evaluation of the cyclic behavior including coupled processes and parameters of the problem remains a task for future work.

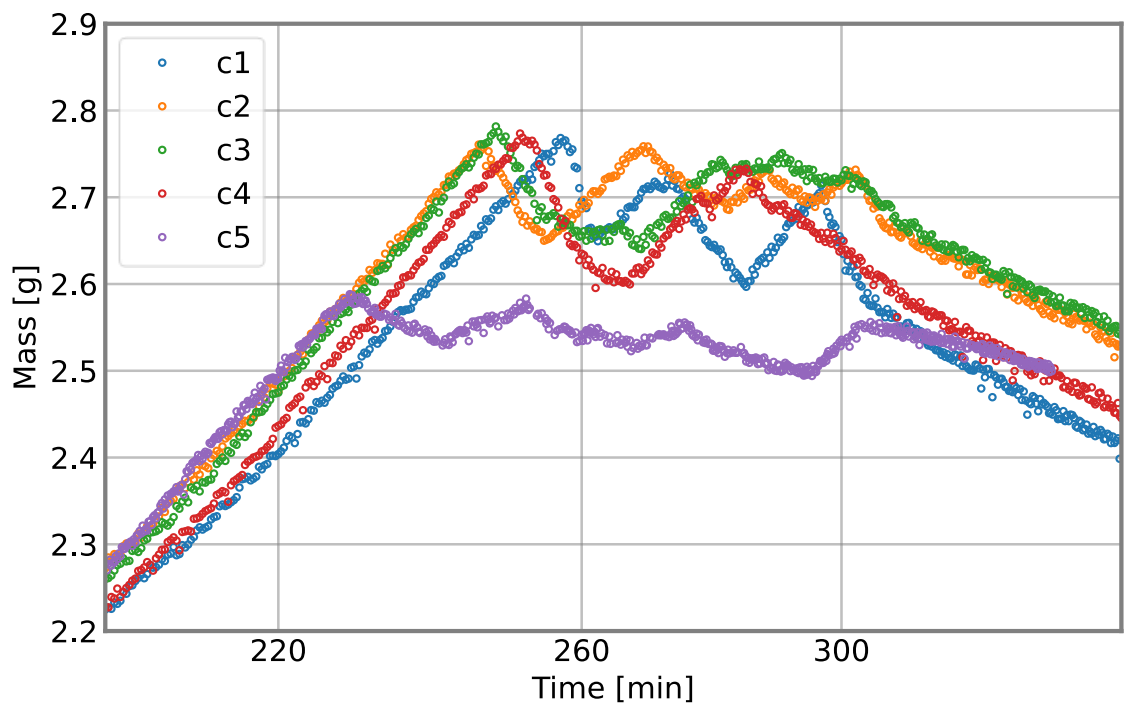

**Figure 11.** – Fluctuations in mass of mobile gas [g] in Box A after initial spilling event. The mass curves all demonstrate oscillations due to recurring spilling events from the anticline to the adjacent fault. For all runs, the maximum mass was observed before the initial gas escape. The lower atmospheric pressure for run C5 (purple circles) results in a lower initial spilling time.

### 4. Concluding remarks

The open-access, high-quality laboratory dataset, accompanied with dedicated analysis tools, represents an asset and opportunity for the carbon storage community to expand the current analysis in future studies. The physical data, describing many of the relevant processes for subsurface carbon storage, may also be used for model validation, comparison, and data-driven forecasts for different stages of a carbon storage operation. Blueprints of the experimental infrastructure enhance reproducibility of scientific research, and enable the porous media community at large to build physical assets and collectively join our efforts.

Our outlook, based on the observations identified in this study, is to probe the origin and premises for establishing non-thermally induced oscillating flows, and to broaden the understanding of at what length scales and to what accuracy multiphase flows in porous media are deterministic.

In conclusion, the observed processes and phenomena qualitatively corroborates the physical understanding and knowledge within the carbon storage community. This supports the assertion that we have a sufficient understanding to claim that industrial carbon storage operations can be conducted in an efficient and safe manner.

### 5. Acknowledgements

The work of JWB is funded in part by the UiB Akademia-project «FracFlow» and the Wintershall DEA-funded project «PoroTwin» MH is funded from Research Council of Norway (RCN) project no. 280341. KE and MH are partly funded by Centre for Sustainable Subsurface Resources, RCN project no. 33184. BB is funded from RCN project no. 324688.

# References

Ayub, M. and Bentsen, R.G. Interfacial viscous coupling: a myth or reality?, J. of Petroleum Science and Engineering (1999) https://doi.org/10.1016/S0920-4105(99)00003-0

Bastesen, E. and Rotevatn, A., Evolution and structural style of relay zones in layered limestone–shale sequences: insights from the Hammam Faraun Fault Block, Suez rift, Egypt, J. of the Geological Society (2012) http://dx.doi.org/10.1144/0016-76492011-100

Both, J.W., Storvik, E., Nordbotten, J.M., Benali, B. DarSIA v1.0, (2023a) https://doi.org/10.5281/zenodo.7515016

Both, J.W., Benali, B., Folvord O., Haugen M., Storvik E., Fernø, M.A., Nordbotten, J.M., Image analysis of the International FluidFlower Benchmark dataset, Zenodo (2023b), https://doi.org/10.5281/zenodo.7515038

Cole J.C. Van De Ven and Mumford, K.G., Intermediate-Scale Laboratory Investigation of Stray Gas Migration Impacts: Transient Gas Flow and Surface Expression, Environ. Sci. Technol (2020) https://doi.org/10.1021/acs.est.0c03530

Eikehaug, K., Haugen, M., Folkvord, O., Benali, B., Bang Larsen, E., Tinkova, A., Rotevatn, A., Nordbotten, J.M., Fernø, M.A. Engineering meter-scale porous media flow experiments for quantitative studies of geological carbon sequestration, TiPM SI (2023a), submitted

Eikehaug, K., Bang Larsen, E. Haugen, M., Folkvord, O., Benali, B., Both, J.W., Nordbotten, J.M., Fernø, M.A. The International FluidFlower benchmark study dataset (2023b) https://:doi.org.10.5281/zenodo.7510589

Elenius, M.T., Nordbotten, J.M., Kalisch, H. Effects of a capillary transition zone on the stability of a diffusive boundary layer, IMA Journal of Applied Mathematics (2012) https://doi.org/10.1093/imamat/hxs054

Flemisch B, Nordbotten JM, Fernø MA, Juanes R, Class H, Delshad M, Doster F, Ennis-King J, Franc J, Geiger S, Gläser D, Green C, Gunning J, Hajibeygi H, Jackson SJ, Jammoul M, Karra S, Li J, Matthäi SK, Miller T, Shao Q, Spurin C, Stauffer P, Tchelepi H, Tian X, Viswanathan H, Voskov D, Wang Y, Wapperom M, Wheeler MF, Wilkins A, Youssef AA, Zhang Z. *The FluidFlower International Benchmark Study: Process, Modeling Results, and Comparison to Experimental Data,* TiPM SI (2023), submitted

Furre, A.K., Eiken, O., Alnes, H., Vevatne, J.N., Kiær, A.F., 20 Years of Monitoring $CO_2$-injection at Sleipner, Energy Procedia (2017) https://doi.org/10.1016/j.egypro.2017.03.1523

Nixon, C.W., Nærland, K., Rotevatn, A., Dimmen, V., Sanderson, D.J., Kristensen, T.B., Connectivity and network development of carbonate-hosted fault damage zones from western Malta, J. of Structural Geology (2020) https://doi.org/10.1016/j.jsg.2020.104212

Geophysical Institute, University of Bergen, https://veret.gfi.uib.no/?action=download

Haugen, M., Saló-Salgado, L., Eikehaug, K., Benali, B., Both, J.W., Storvik, E., Folkvord, O., Juanes, R., Nordbotten, J.M., Fernø, M.A., Physical variability in meter-scale laboratory $CO_2$ injections in faulted geometries, TiPM SI (2023), submitted

Karstens, J., Ahmed, W., Berndt, C. and Class, H., Focused fluid flow and the sub-seabed storage of $CO_2$: Evaluating the leakage potential of seismic chimney structures for the Sleipner $CO_2$ storage operation, Marine and Petroleum Geology (2017) https://doi.org/10.1016/j.marpetgeo.2017.08.003

Karstens, J. and Berndt, C., Seismic chimneys in the Southern Viking Graben–Implications for palaeo fluid migration and overpressure evolution, Earth and Planetary Science Letters (2015) https://doi.org/10.1016/j.epsl.2014.12.017

Kovscek, A.R., Nordbotten, J.M., Fernø, M.A., Scaling up FluidFlower results for carbon dioxide storage in geological media, TiPM SI 2023, submitted.

Lemmon, E.W., Bell, I.H., Huber, M.L., McLinden, M.O., *Thermophysical Properties of Fluid Systems*, NIST Chemistry WebBook, NIST Standard Reference Database Number 69, Eds. P.J. Linstrom and W.G. Mallard, National Institute of Standards and Technology, Gaithersburg MD, 20899, https://doi.org/10.18434/T4D303, (retrieved September 2, 2022).

Nordbotten JM, Fernø MA, Flemisch B, Juanes R, Jørgensen M (2022). Final Benchmark Description: FluidFlower International Benchmark Study. Zenodo. https://doi.org/10.5281/zenodo.6807102

Nordbotten, J.M., Benali, B., Both, J.W., Brattekås, B., Storvik, E., Fernø, M.A.; Two-scale image processing for porous media, TiPM SI (2023), submitted

Ogata, K., Senger, K., Braathen, A., Tveranger, J., Fracture corridors as seal-bypass systems in siliciclastic reservoir-cap rock successions: Field-based insights from the Jurassic Entrada Formation (SE Utah, USA), J. of Structural Geology (2014) https://doi.org/10.1016/j.jsg.2014.05.005

Pau, G.S.H., Bell, J.B., Pruess, K., Almgren, A.S., Lijewski, M.J., and Zhang, K., High-resolution simulation and characterization of density-driven flow in $CO_2$ storage in saline aquifers, Adv. Water Resour., (2010), https://doi.org/10.1016/j.advwatres.2010.01.009

Pruess, K., Numerical studies of fluid leakage from a geologic disposal reservoir for $CO_2$ show self-limiting feedback between fluid flow and heat transfer, Geophys. Res. Lett. (2005) http://doi.org.10.1029/2005GL023250.

Riaz, A., Hesse, M., Tchelepi, H., Orr, F. Onset of convection in a gravitationally unstable diffusive boundary layer in porous media. Journal of Fluid Mechanics, 548, 87-111. (2006) http://doi.org.10.1017/S0022112005007494

Rotevatn, A., Buckley, S.J., Howell, J.A. Fossen, H. Overlapping faults and their effect on fluid flow in different reservoir types: A LIDAR-based outcrop modeling and flow simulation study, AAPG Bulletin (2009) https://doi.org/10.1306/09300807092

Sharma, M. Alcorn, Z.P., Fredriksen, S.B., Rognmo, A.U., Fernø, M.A., Skjæveland, S.M., Graue, A. Model calibration for forecasting $CO_2$-foam enhanced oil recovery field pilot performance in a carbonate reservoir, Petroleum Geoscience (2020) https://doi.org/10.1144/petgeo2019-093

Shipton, Z. K., J. P. Evans, D. Kirschner, P. T. Kolesar, A. P. Williams, and J. Heath, Analysis of $CO_2$ leakage through 'low-permeability' faults from natural reservoirs in the Colorado Plateau, east-central Utah, Geological Storage of Carbon Dioxide (2004), edited by S. J. Baines and R. H. Worden, Geol. Soc. London Spec. Publ., 233, 43–58

**Supplementary materials**

**SM 1 – Atmospheric pressure and volume variations between runs C1-C5**

*Atmospheric pressure for all runs C1 to C5.* The atmospheric pressure in the laboratory varied during the experimental campaign (executed between November 24th 2021 and Jan 14th 2022). The variations in atmospheric pressure in the lab used historical metrological pressure data recorded approximately 100 meters from the FluidFlower rig (see **Figure SM1**).

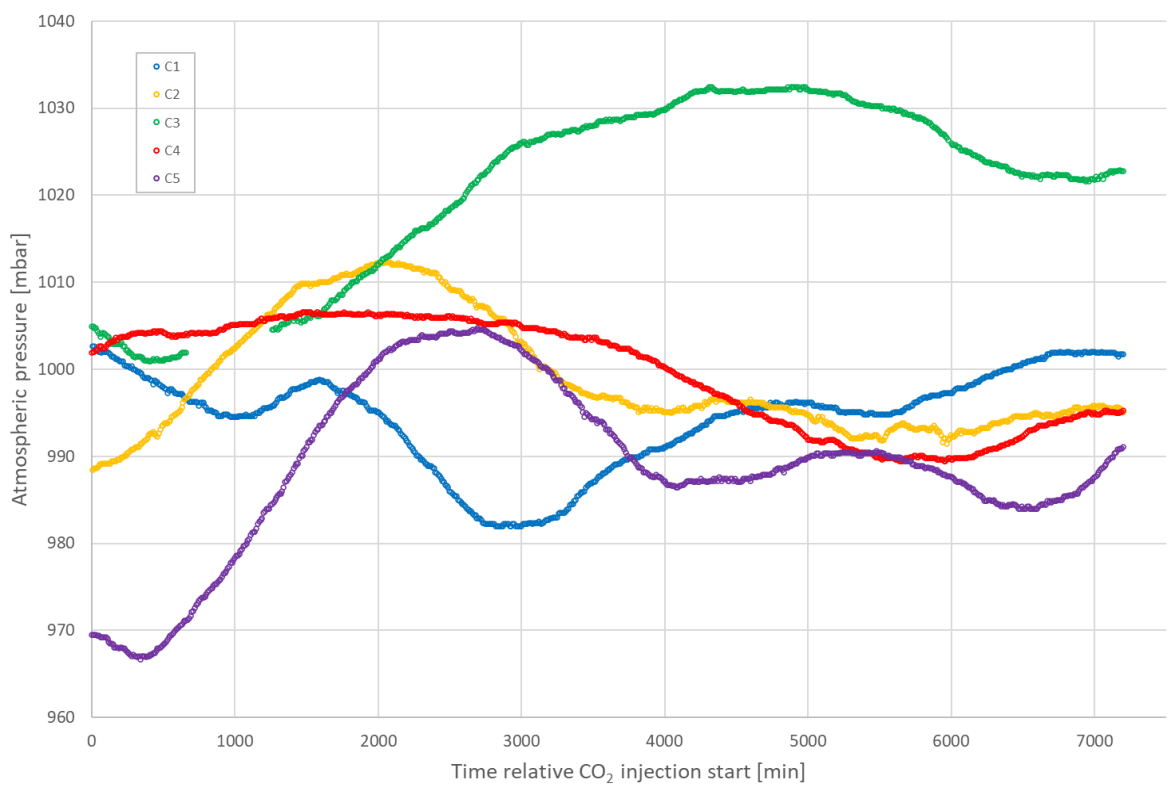

**Figure SM.1** - The atmospheric pressure recorded for each of the five runs C1-C5 [Geophysical Institute, 2022] during the experimental campaign (November 2021 to January 2022).

The average atmospheric pressure during the injection period (see **Table SM.1**) determined the volume of injected CO2 and influenced some parameters in Table 3 and degree of physical reproducibility.

**Table SM.1**. Overview of atmospheric pressure variations between runs C1-C5

| Cycle | Atmospheric pressure [mbar] | | |
|---|---|---|---|
| | <$CO_2$ injection> ± σ | Max | Min |
| 1 | 1001 ± 1 | 1003 | 982 |
| 2 | 990 ± 1 | 1012 | 990 |
| 3 | 1003 ± 1 | 1032 | 1003 |
| 4 | 1003 ± 1 | 1007 | 1003 |
| 5 | 968 ± 1 | 1005 | 967 |

## SM 2 – Depth map and sand sedimentation process

The depth of the porous media in the rig was measured after all CO2 injection experiments and reported here.

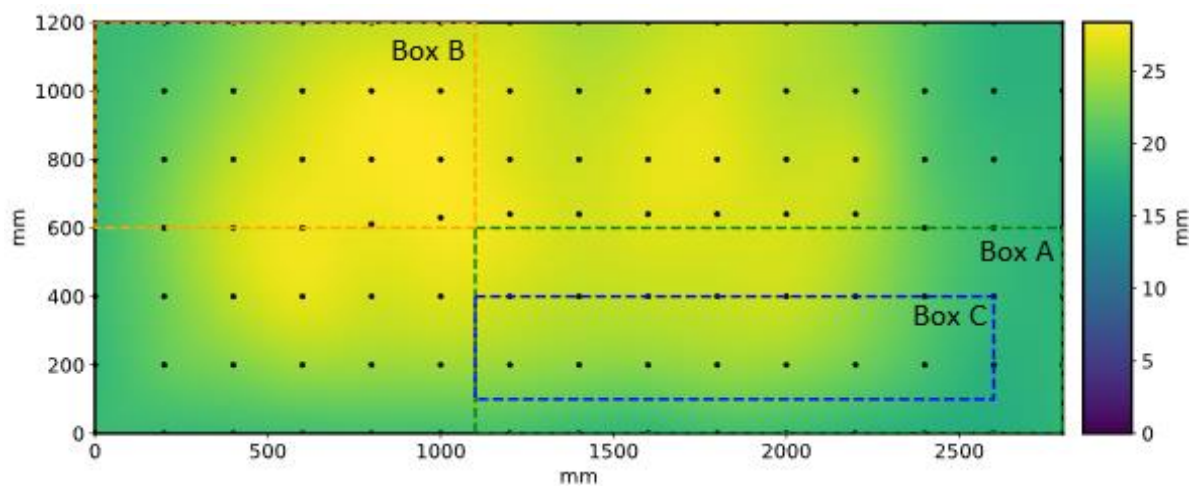

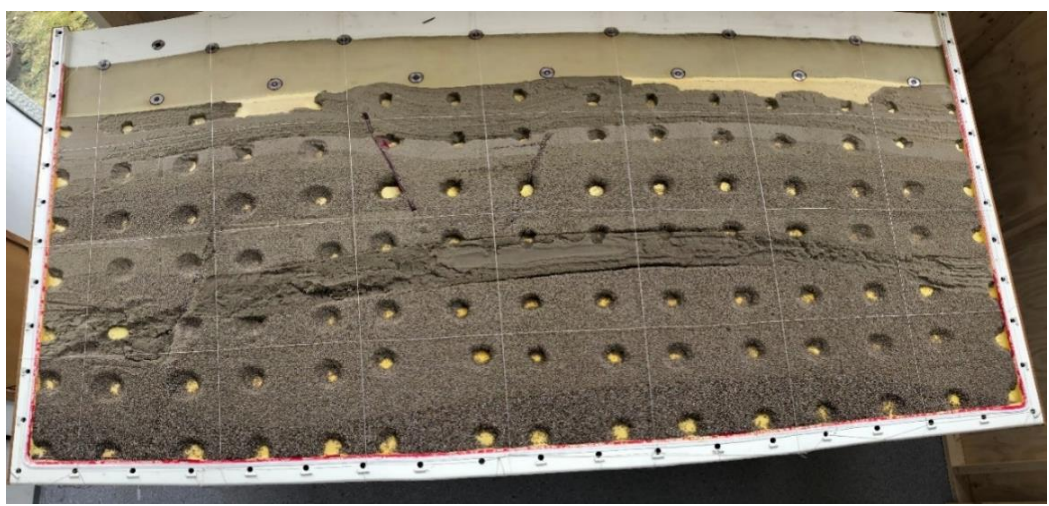

**Figure SM.2 -** Photograph of the rig (positioned horizontally) after all runs and the front panel removed. Sand was carefully removed in isolated regions of the geometry, and the sand depth was measured in each location (black dots in depth map, cf. **Table SM.2**)

**Table SM.2.** Measured sand depth (in mm) at isolated points (cf. Figure SM2) after all runs. x- and y-coordinates are in mm from lower left corner.

| | | | | | | | | | | | | | | |
|---|---|---|---|---|---|---|---|---|---|---|---|---|---|---|
| **1200** | 19.2 | 20 | 22.7 | 24.8 | 25.7 | 25.9 | 26 | 24.5 | 25.3 | 26.2 | 25.3 | 23.7 | 20.9 | 18.4 | 18.5 |
| **1000** | 19.2 | 20.9 | 23.8 | 26.3 | 27.8 | 27.7 | 26.9 | 25.5 | 26.5 | 26.9 | 25.3 | 24.5 | 20.9 | 18.4 | 18.5 |
| **800** | 19.2 | 22.1 | 25.5 | 27.3 | 28.1 | 28.3 | 27 | 26.2 | 27.2 | 27.4 | 26 | 26 | 21.2 | 19.2 | 18.5 |
| **600** | 19.2 | 23 | 26.5 | 27.8 | *27.5* | **28.1** | ***27.6*** | ***26.4*** | ***27.1*** | ***26.9*** | ***26.6*** | ***25.4*** | 21.7 | 19.4 | 18.2 |
| **400** | 19.4 | 22.3 | 25.6 | 27.4 | 26.4 | 26.9 | 26.3 | 25.6 | 25.8 | 25.9 | 26.1 | 24.1 | 21.2 | 19 | 18.5 |
| **200** | 18.9 | 21.6 | 22.9 | 24.3 | 24.2 | 24.3 | 24.1 | 23.6 | 23.5 | 24.1 | 23.8 | 22.3 | 20.2 | 18.2 | 18.7 |
| **0** | 19.2 | 19.2 | 19.4 | 19.7 | 19.9 | 19.8 | 19.8 | 19 | 18.6 | 19.6 | 19.3 | 18.5 | 17.9 | 17.5 | 18.5 |
| **H/L** | **0** | **200** | **400** | **600** | **800** | **1000** | **1200** | **1400** | **1600** | **1800** | **2000** | **2200** | **2400** | **2600** | **2800** |

Note: Depths on row '600' were as measured at 610 mm (italic), 630 mm (bold) and 640 mm (bold + italic)

**SM 3 – Image analysis assumptions**

A range of assumptions (I. to V.) was needed to quantitatively describe the observed multiphase flow phenomena during repeated $CO_2$ injections in the physical flow rig. Based on these assumptions, a geometric separation of the formation water from any $CO_2$ in the system and of the gaseous $CO_2$ from the $CO_2$-saturated water is sufficient.

I. *we assume that gas-filled regions are 100% saturated with the gas ($CO_2$)*
Note: we physically understand that there is a non-zero irreducible water saturation ($S_{wi}$) present in parts of the gas saturated regions. The regions with free gas (predominantly under the lower anticline seal) have a non-zero water content due to *i*.) solubility of water in $CO_2$ (the gas travels some distance from injection well [9,3] through a water-saturated region before it accumulates) and *ii*.) an irreducible water saturation may be present, determined by the capillary drainage curve and the 'wetness' of the gas displacing water. Neither i) nor ii) can accurately be quantified in our setup, and we make a general assumption that $S_{wi} = 0$ in gas-filled regions.

II. *we assume a constant $CO_2$ concentration in the $CO_2$-saturated water*
Note: This is a data limitation rather than an assumption, as we expect a $CO_2$ concentration gradient in $CO_2$-saturated water depending on the distance from the injector and time; in particular, it is expected based on numerical simulations that the concentration in general will be variable within in each gravitational finger (Pau et al 2010). In most experimental runs, however, we observed mineralization of methyl red from the blend of pH sensitive dyes in the formation water. Hence, the anticipated relationship between color change and $CO_2$ concentration (due to pH changes) could not be quantified.

III. *we do not account for the dynamics of the gas partitioning in the gas accumulation*
Note: The formation water was in equilibrium with the atmosphere and contained dissolved atmospheric gases (predominantly nitrogen and oxygen). The $CO_2$ mass transfer from gaseous phase to the formation water releases nitrogen and oxygen into the gaseous phase. Hence, over time the gaseous phase in the system becomes deprived of $CO_2$, with reduced solubility in water. We estimate the amount of atmospheric gases in observed gaseous phase after the experiment.

IV. *we can accurately calculate the volume of $CO_2$ injected during each run*
Note: The mass flow controller ensures that the mass of $CO_2$ injected in each run is identical (within instrument uncertainty). The volume of the injected $CO_2$ is, however, influenced by the system pressure and temperature. The logged atmospheric pressure during the injection phase (average for entire injection period) is used, together with the hydrostatic pressure, to calculate the injected $CO_2$ volume for each run, assuming uniform and constant temperature.

V. *we have accurate information about porosity and depth*
Note: the experimentally measured porosity values for each sand type represent the porosity in the FluidFlower rig, and we assume that the porosity is uniform in each layer. The variable depth measured after completion of all runs is representative for all runs, with radial basis function (RBF) interpolation between measured locations. The depth and porosity are used to calculate the pore volume in each sand layer.

**SM 4 – Fluid injection protocols**

Here we briefly describe the two main gaseous and aqueous injection protocols used during the benchmark study.

*Gaseous phase injection protocol*

Gaseous $CO_2$ was injected using a mass flow controller (EL-FLOW Prestige FG-201CV 0-10 ml$_s$/min, BronkHorst) with a constant mass per time injection rate equal to 10.000 ± 0.0002 ml$_s$/min using the following injection protocol (cf. **Figure 1** for injection well positions):

1. Inject $CO_2$ in well [9,3]; duration 300 minutes
2. Inject $CO_2$ in well [17,7] with 135 minutes delay relative to well [9,3]; duration 165 minutes
3. Stop $CO_2$ injections in wells [9,3] and [17,7] simultaneously
4. Monitor $CO_2$ flow, convection, trapping and dissolution for 120 hours (five days)

Note: the injection time in well [9,3] is 304.5 min due to ramping: increased from 1 to 10 ml$_s$/min over the first 4.5 min (1 ml$_s$/min increments every 30 sec). At the end of $CO_2$ injection, the delivery rate likewise decreased. Hence, the total injection time of 304.5 min with up/ down ramping equals 300 min with constant 10 ml$_s$/min. A similar approach was used for $CO_2$ injection in well [17,7], where injection time with ramping (169.5 min) equals 165 min with 10 ml$_s$/min delivery rate.

The injected $CO_2$ mass is the same for each run: well [15,5] = 5.5167 g; well [17,11] = 2.94224 g; total $CO_2$ mass injected = 8.45894 g. The masses are calculated multiplying the injection time (min), delivery rate (ml$_s$/min) and density (g/ml), where $\rho_{CO2}$ at standard conditions is 0.0018389 g/ml [Lemmon et al. 2022]. ml$_s$ equals ml/min at standard conditions: T = 20 °C and P = 1013 mbar.

*Aqueous phase injection protocol*

The protocol below describes the preparatory steps before $CO_2$ injection and how to reset the fluids in the geometry for repeated runs. All aqueous phases were delivered to the flow rig with a piston pump (QX-1500, Chandler Engineering). The following steps were performed:

1. *Prepare for $CO_2$ injection and saturate the porous media*: inject formation water with 50 ml/min injection rate through the seven ports in the technical row at the bottom of the rig. Port positions (open/close) was varied sequentially to achieve an effective miscible displacement process.
2. *Maintain fixed water table height*: recirculate formation water (injection rate 50 ml/min) in the top technical row during $CO_2$ injection and dissolution
3. *Remove gas and $CO_2$-saturated water after $CO_2$ experiment:* inject lye solution with 50 ml/min through the seven ports in the bottom technical row.
4. Repeat steps 1-3.

**SM 5 Physical variability**

Supporting information regarding the degree of physical overlap and reproducibility.

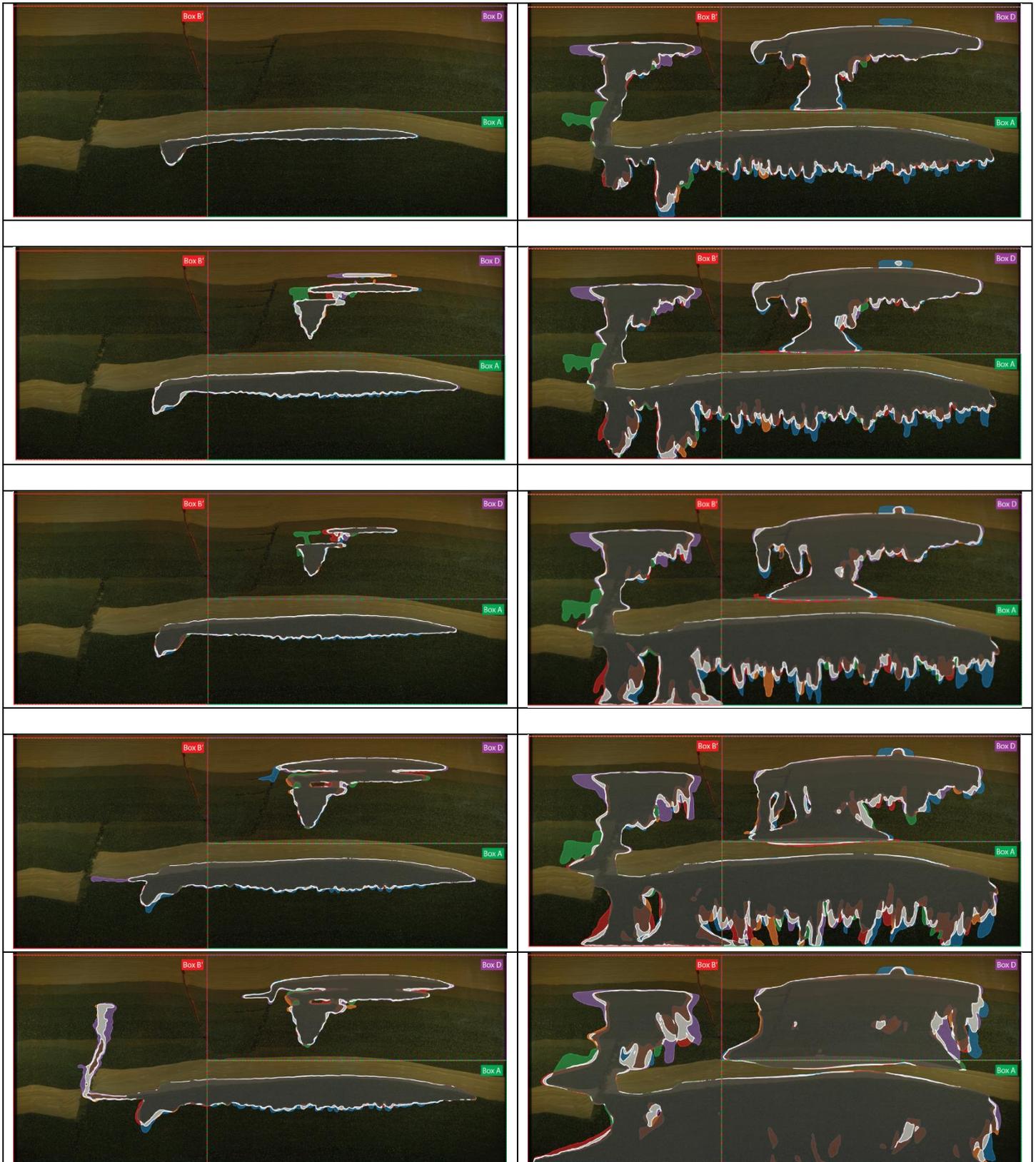

**Figure SM.3** - Spatially distributed overlap for all runs, with the following color scheme: gray (overlap C2+C3+C4); blue (unique C1); orange (unique C2); green (unique C3); red (unique C4); purple (C5 unique); brown (combinations all runs with at least one of C2, C3 or C4), white (other combinations). <u>Left column</u>: time steps describing the *pre-spilling interval* at 60 min (top); 155 min; 230 min; 255 min; 300 min (bottom). <u>Right column</u>: time steps describing the *gravitational fingers interval* at 10 hours (top); 14 hours; 20 hours; 30 hours; and *dissolution-driven flow interval* at 66 hours (bottom). Intervals are defined in **chapter 3.2.**

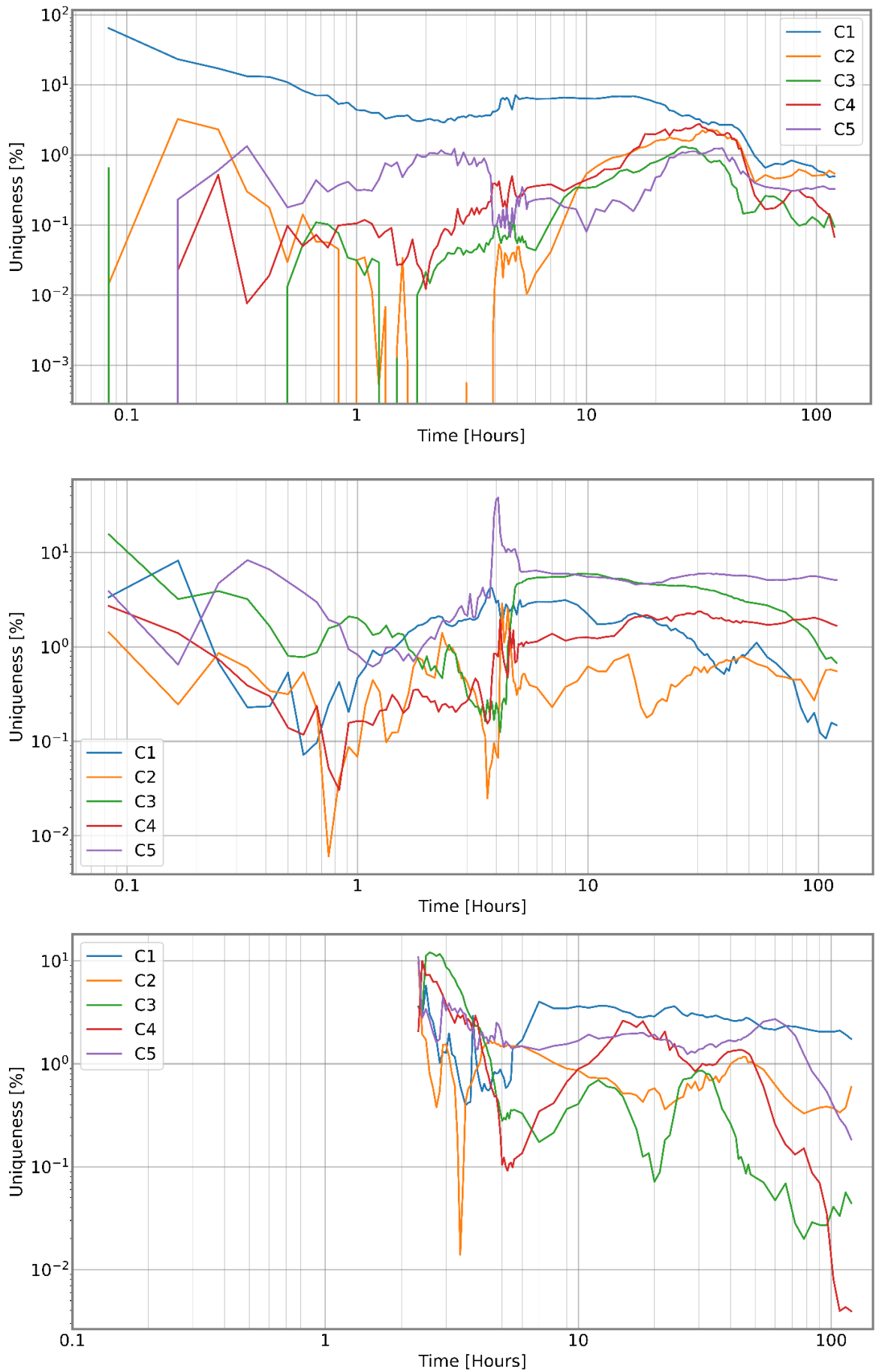

**Figure SM.4** - Uniqueness of each run C1-C5 in Box A (top), Box B' (middle) and Box D (bottom). Note that all axes are logarithmic.

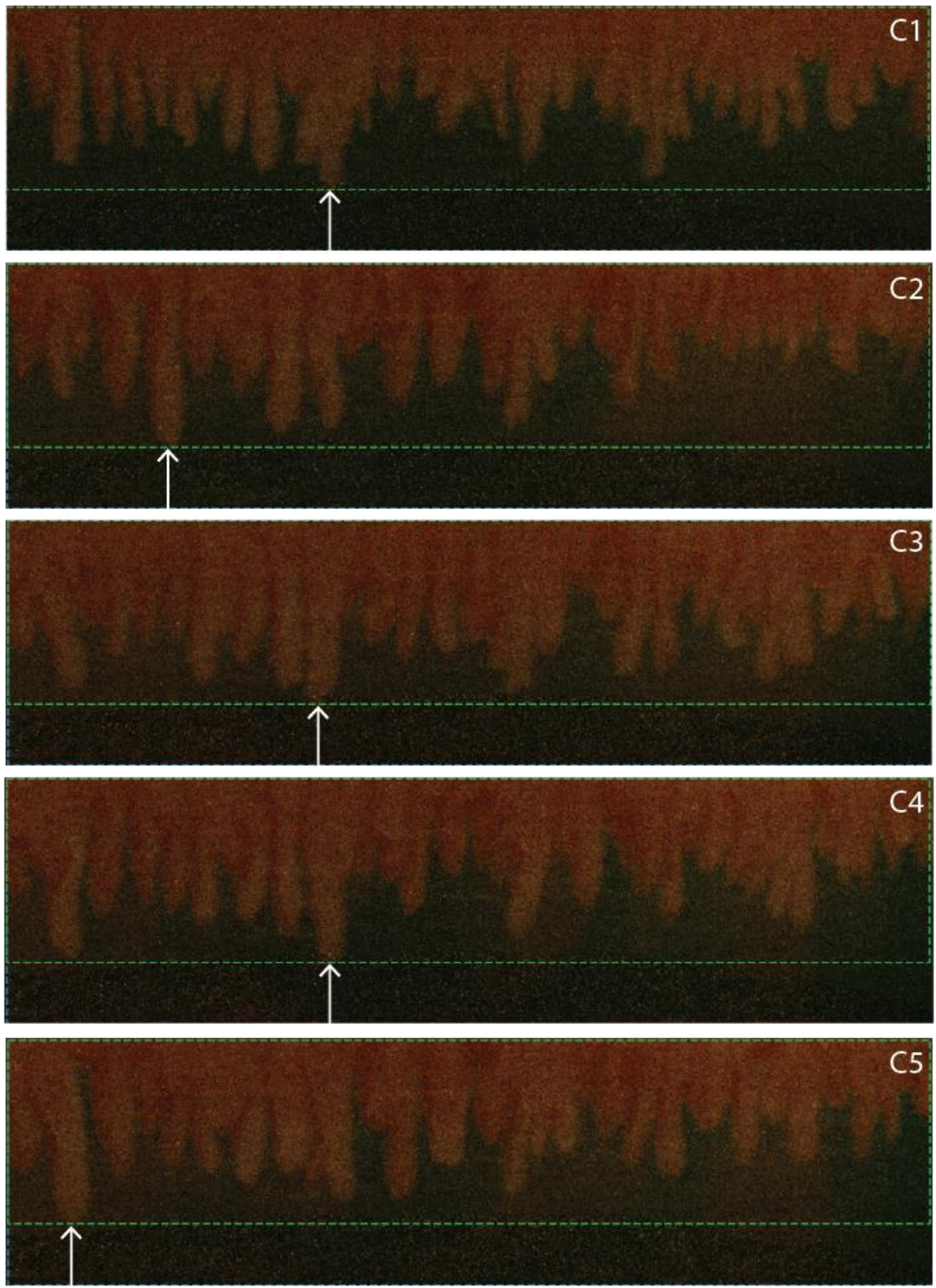

**Figurer SM.5** - Images of gravitational fingers for each run C1 to C5 when the first finger exist the lower boundary of Box C (green dashed line). For run C1, C3 and C4 it was the same finger that first reached the lower boundary, and overall the fingers appear reproducibile. Finger width is somewhat smaller for run C1, a combined effect for the earlier exist (cf. **Table SM.3**) and inconcistent water chemistry.

**Table SM.3. Metric of finger characteristics for runs C1 to C5**

|  | C1 | C2 | C3 | C4 | C5 | Average |
|---|---|---|---|---|---|---|
| First finger enters Box C [min] | 105 | 115 | 120 | 110 | 110 |  |
| First finger exists Box C [h] | 14 | 17 | 20 | 19 | 20 |  |
| Contour length when first finger exists [m] | 4.17 | 3.96 | 3.52 | 3.41 | 4.02 |  |
| Maximum number of fingers | 35 | 34 | 33 | 32 | 35 | 34 ± 1 |

**SM 6 Sand sedimentation**

Using the image analysis capabilities of DarSIA, it is possible to detect minute structural changes the sand facies over time, resulting in a local deformation map matching the pore spaces of two images. The multi-level feature detection methodology, as detailed in [Nordbotten et al. 2023], allows for an accurate analysis on the resolution of grains. Using this methodology, we can investigate sand sedimentation occurring over the course of the various experimental runs, to assess whether the geometry has sedimented to a final configuration modulo insignificant variations. This is particularly important for discussing physical variability of the different experimental runs.

To assess the significance of sand sedimentation in between two operational runs and over the experimental study, we study the effective displacement between two runs (C1 and C2) and total displacement (between C1 and C5) cf. **Figure SM.5**. The maximal observed sedimentation occurs on the order of 1.7 mm and 5.5 mm for the two respective options. We conclude that the changes in the geometry are insignificant in terms of their effect on the flow behavior.

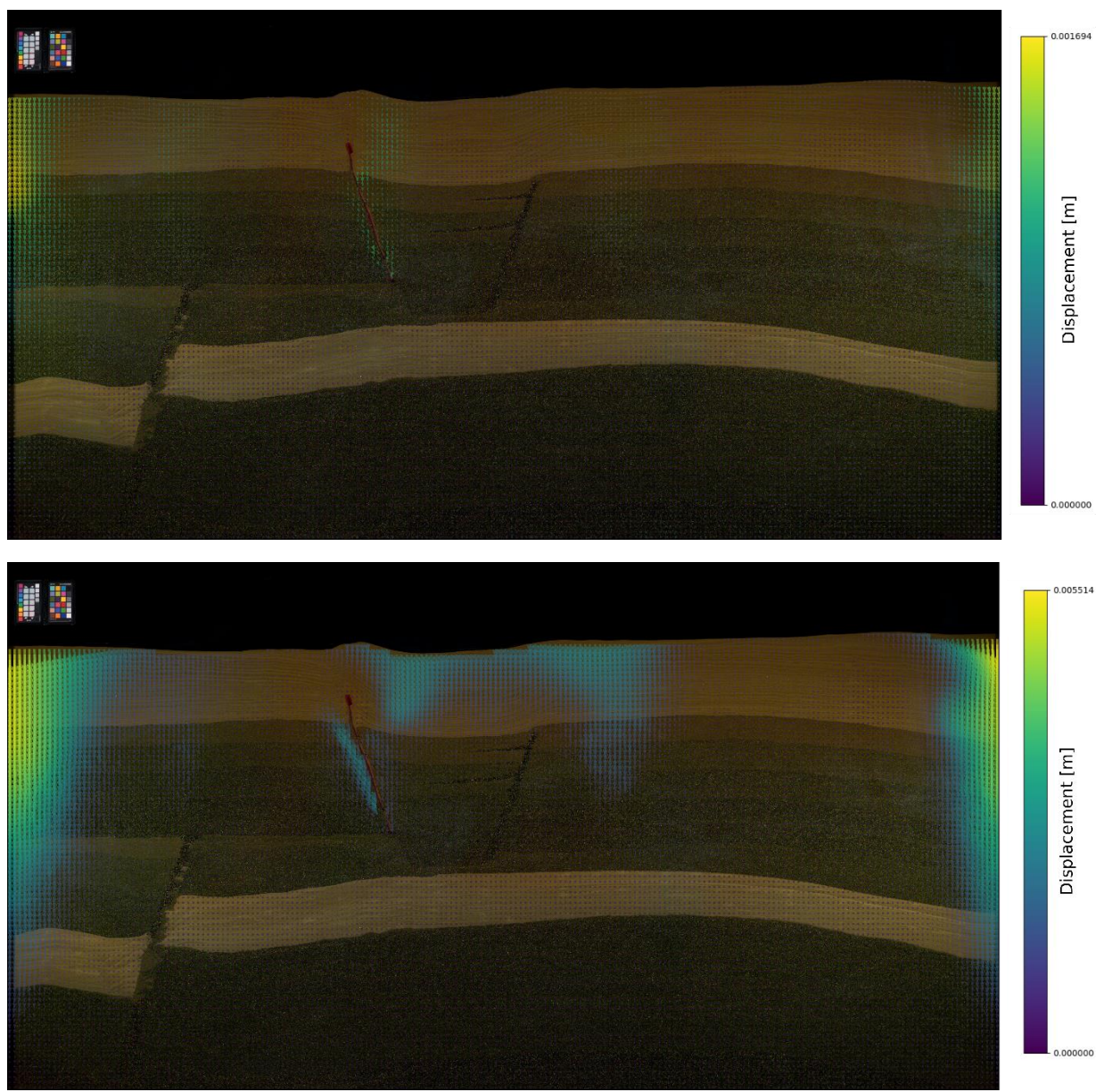

**Figure SM.6** - Sand sedimentation during CO2 injections. Top: Deformation aligning pore spaces the geometries at injection start of runs C1 and C2. The effective sedimentation is most significant along the sides of the geometry as well as along the sealed fault, with a maximal displacement of 1.7 mm. Bottom: Deformation aligning pore spaces the geometries at injection start of runs C1 and C5, illustrating the overall sedimentation taking place over the course of the entire experimental study. The maximal displacement along the sides reaches values of 5.5 mm.